\documentclass[prb,aps,twocolumn,superscriptaddress,10pt]{revtex4-2}

\usepackage{graphicx}

\usepackage{mathtools}

\usepackage{amsmath}

\usepackage{amssymb}

\usepackage{bm}

\usepackage[dvipsnames]{xcolor}

\usepackage[
	colorlinks=true,
	linkcolor=blue,
	anchorcolor=blue,
	citecolor=blue,
	urlcolor=blue
]{hyperref}

	\newcommand{\ket}[1]{|{#1}\rangle}
	
	\newcommand{\vev}[1]{\langle{#1}\rangle}
	
	\newcommand{\comm}[2]{[{#1},{#2}]}
	
	\newcommand{\acomm}[2]{\{{#1},{#2}\}}

	\DeclareMathOperator{\tr}{tr}
	
	\DeclareMathOperator{\sgn}{sgn}
	
	\DeclareMathOperator{\hc}{H.c.}

	\DeclareMathOperator{\cyclic}{cyclic\ perm}
	
\begin{document}

%%%%%%%%%%%%%%%%%%%%%%%%%%%%%%%%%%%%%%%%%%%%%%%%%%%%%%%%%%

\title{Chiral fixed point in a junction of critical spin-1 chains}

\author{Hernan B. Xavier}
\affiliation{Departamento de F\'isica Te\'orica e Experimental, Universidade Federal do Rio Grande do Norte, 59072-970 Natal-RN, Brazil}

\author{Rodrigo G. Pereira}
\affiliation{Departamento de F\'isica Te\'orica e Experimental, Universidade Federal do Rio Grande do Norte, 59072-970 Natal-RN, Brazil}
\affiliation{International Institute of Physics, Universidade Federal do Rio Grande do Norte, 59078-970 Natal-RN, Brazil}

\date{\today}

\begin{abstract}

%	motivação
Junctions of one-dimensional systems are of great interest to the development of synthetic materials  that harbor topological phases.
%	métodos e resultados
We study a junction of three gapless spin-1 chains described by the $\mathrm{SU(2)}_{2}$ Wess-Zumino-Wittten model  and coupled by   exchange and chiral three-spin   interactions. We show that a chiral fixed point appears as a special point on the transition line   separating two regimes described by open boundary conditions, corresponding to decoupled chains and the formation of a boundary spin singlet state. Along this transition line, the junction behaves as a tunable spin circulator as the spin conductance varies continuously with the coupling constant of a marginal boundary operator. 
%	conclusões
Since the spectrum of the junction contains fractional excitations such as Majorana fermions, in this paper, we set the stage for network constructions of non-Abelian  chiral  spin liquids.

\end{abstract}

\maketitle

%%%%%%%%%%%%%%%%%%%%%%%%%%%%%%%%%%%%%%%%%%%%%%%%%%%%%%%%%%

\section{Introduction}

Junctions of multiple one-dimensional (1D) electronic systems, such as quantum wires and spin chains, are of great relevance   to technological applications because they constitute basic   elements in the architecture of quantum devices \cite{Alicea2011,Qiu2020}. Networks of  1D conducting channels also provide versatile  platforms to simulate exotic phases of matter \cite{Medina2013,SanJose2013,Ferraz2019,Lee2021,Chou2021}. On the  theoretical side, junctions of 1D systems offer an amenable,  yet nontrivial, playground to explore fascinating phenomena associated with strong correlations. Their transport properties can be characterized by renormalization group (RG) fixed points of the low-energy theory that correspond to conformally invariant boundary conditions of the collective charge or spin modes   \cite{Cardy1984,Cardy1991}. This approach has been applied to 
several quantum impurity problems, a prominent example being  the multichannel Kondo model \cite{Affleck1991,Affleck1993}.

Different boundary fixed points can be reached  depending  on boundary couplings as well as bulk interactions \cite{Vojta2006}. In the context of quantum wires,   Kane and Fisher \cite{Kane1992a,Kane1992b} showed that electrons are fully transmitted through a weak link      if the electron-electron interactions are attractive and perfectly reflected otherwise. In another influential work, Chamon \textit{et al.} \cite{Chamon2003}  mapped out the phase diagram of a Y junction with an enclosed magnetic flux. Remarkably, they identified  a chiral fixed point in which the sign of the magnetic flux controls the asymmetry  in the current flow \cite{Barnabe2005,Oshikawa2006}.  
 
In parallel to these developments, many spin chain models have been studied as well \cite{Eggert1992,Bayat2012,Bayat2014,Tsvelik2013,Giuliano2016}. In particular, Buccheri \emph{et al.}  \cite{Buccheri2018,Buccheri2019} showed that a    Y junction of spin-1/2 antiferromagnetic Heisenberg chains  features an unstable chiral fixed point that can be reached by fine tuning a single coupling constant, namely, the strength of a boundary  three-spin interaction  that breaks time reversal   symmetry. The intermediate-coupling  chiral fixed point appears at the transition between two stable fixed points characterized by decoupled   chains and by a spin-chain analog of the three-channel Kondo model.  Precisely at the chiral fixed point,  the junction behaves as an ideal quantum spin circulator. Since each spin-1/2 chain is described by an $\mathrm{SU(2)}_{1}$ Wess-Zumino-Witten (WZW) model \cite{Gogolin1998}, the chiral junction also provides a starting point for network constructions of Abelian chiral spin liquids in higher dimensions \cite{Ferraz2019}. An open question is whether this chiral fixed point can be generalized to higher-level $\mathrm{SU(2)}_{k}$ WZW models, known to describe critical points of isotropic   chains with spin $S\geq1$  \cite{Affleck1987,Nielsen2011,Michaud2012,Thomale2012,Michaud2013}. Such a generalization  would pave the way for realizing  more exotic chiral spin liquid states with non-Abelian spinons \cite{Greiter2009,Yao2011} in networks of gapless spin-$S$ chains.

In this paper we pursue this generalization by  studying a Y junction of critical spin-1 chains whose continuum limit is described by three copies of the $\mathrm{SU(2)}_{2}$ WZW model. Each copy  has central charge $c=3/2$ and can be represented in terms of three critical  Ising models \cite{Tsvelik1990,Allen2000}. As a consequence, the excitation spectrum  contains emergent Majorana fermions.  To construct the junction, we coupled the spin chains  by  exchange  and    three-spin interactions  among their end spins. Our main goal is to locate the  chiral fixed point in the boundary phase diagram by inspecting   its instabilities within the effective field theory. Based on an analysis of the effects of  relevant and marginal perturbations, combined with the picture for weak and strong coupling limits of the lattice model, we argue that the chiral fixed point  lies on the transition line  separating the regime where the chains are trivially decoupled from the regime where the boundary spins form a singlet state.  Along this line, we compute  spin transport properties  by solving a scattering problem for non-interacting chiral   fermions. We find that   the  spin conductance tensor can be partially  asymmetric, with a maximal asymmetry at the chiral fixed point. 

This paper is organized as follows. Section \ref{sec:model} starts with a review of  the low-energy   theory for the critical spin-1 chain that models the junction legs. Having set up the notation, we turn to the effective field theory for the junction and discuss the open-boundary  fixed point   governing the weak-coupling regime. In Sec. \ref{sec:chiral-FP} we begin to characterize the chiral fixed point of our junction by calculating the long-distance decay of the three-spin correlation. We also address the boundary operators that render  the chiral fixed point unstable. This leads us to the analysis of Sec. \ref{sec:relevant}, where we examine the relevant and marginal perturbations. We find that the relevant operator controls the transition between two phases with open boundary conditions, while  the marginal perturbation by itself affects the spin conductance of the junction. We argue that the  transition line  described by the  marginal deformation of the chiral fixed point   terminates at the point where  an $S=1$ boundary bound state is formed and must be screened by the Kondo effect.  In Sec. \ref{sec:conclusions}, we draw our conclusions and point out some future directions. For self-containing purposes, we also include two appendices. Appendix \ref{app:bosonization} summarizes our conventions for the critical theory of the Ising model and details the bosonization scheme used  in the main text. Finally, Appendix \ref{app:bound-state} shows that a lattice version of the low-energy problem of Sec. \ref{sec:bound}   supports our claim of  a boundary bound state  in the strong coupling limit.

%%%%%%%%%%%%%%%%%%%%%%%%%%%%%%%%%%%%%%%%%%%%%%%%%%%%%%%%%%

\section{Spin model}\label{sec:model}

In this section, we define the model of our junction. We begin with a short review of the $\mathrm{SU(2)}_{2}$ WZW conformal field theory for a critical spin-1 chain. We then introduce the Hamiltonian for the junction, discuss  the role of   the boundary interactions, and examine the fixed point with open boundary conditions as a training example.

%%%%%%%%%%%%%%%%%%%%%%%%%%%%%%%%%%%%%%%%%%%%%%%%%%%%%%%%%%

\subsection{The critical spin-1 chain}

The $\mathrm{SU(2)}_{2}$ WZW universality class arises as  one of the generic possibilities for the transition from the Haldane phase to the dimerized phase in spin-1 chains  \cite{Chepiga2016b}. For concreteness, we consider the bilinear-biquadratic spin-1 chain, with Hamiltonian
\begin{equation}\label{eq:spin-1-chain}
H=J\sum_{j}\Big[(\mathbf{S}_{j}\cdot\mathbf{S}_{j+1})-b(\mathbf{S}_{j}\cdot\mathbf{S}_{j+1})^{2}\Big],
\end{equation}
where $\mathbf{S}_{j}$ are spin-1 operators, $J>0$ is the antiferromagnetic exchange coupling, and the  dimensionless  parameter $b$ controls the relative strength of the biquadratic interaction. This model can be realized with spin-1 bosons in optical lattices \cite{Yip2003,Garcia2004}. The ground-state phase diagram   is well known \cite{Affleck1986,Lauchi2006}.  If we start from   the Heisenberg chain with $b=0$ and increase $b$,  the  transition    to the dimerized phase occurs at the Takhtajan-Babujian point $b=1$, at which the spin Hamiltonian is Bethe ansatz integrable   \cite{Takhtajan1982,Babujian1982}.   The effective Hamiltonian for this critical point has the form
\begin{equation}
H=\int dx \bigg[\frac{\pi v}{2}\big( \mathbf{J}^{2}+\bar{\mathbf{J}}^{2}\big)-2\pi v g \mathbf{J}\cdot\bar{\mathbf{J}}\bigg],
\end{equation}
where $v\sim J$ is the spin velocity, and $\mathbf{J}$ and $\bar{\mathbf{J}}$ are the left- and right-moving spin currents which obey the $\mathrm{SU(2)}_{2}$ Kac-Moody algebra \cite{Gogolin1998}. The dimensionless coupling $g>0$ is marginally irrelevant, producing logarithmic corrections to correlation functions \cite{Affleck1989}. We can tune $g$ to zero by adding   next-nearest-neighbor interactions to the bilinear-biquadratic chain \cite{Chepiga2016a}. To simplify matters and focus on the essential physics, hereafter, we  neglect the marginal bulk interaction.

The $\mathrm{SU(2)_{2}}$ WZW model is very special because it can be expressed as a theory of three critical Ising models  \cite{Tsvelik1990,Gogolin1998,Allen2000}. This means all local operators of the theory can be written as products of Ising operators  labeled by $a\in\{1,2,3\}$. In particular, the components of the spin currents take the form 
\begin{equation}\label{eq:spin-currents}
J^{a}=-\frac{i}{2}\epsilon^{abc}\xi^{b}\xi^{c},\qquad
\bar{J}^{a}=-\frac{i}{2}\epsilon^{abc}\bar{\xi}^{b}\bar{\xi}^{c},
\end{equation}
where $\xi^{a}$ and $\bar{\xi}^{a}$ are the chiral Majorana  fermions associated with each Ising model, and $\epsilon^{abc}$ is the Levi-Civita symbol. These currents represent the smooth part of the continuum representation for the spin operator \cite{Affleck1987}, which reads
\begin{equation}\label{eq:spin-representation}
\mathbf{S}_{j}\sim\mathbf{J}(x)+\mathbf{\bar{J}}(x)+(-1)^{j}\mathbf{n}(x)
\end{equation}
with $x=ja_{0}$, $a_{0}$ the lattice constant. The staggered part of the spin operator is defined in terms of the $2\times2$   matrix field $\Phi^{(1/2)}$ as
\begin{equation}\label{stagg}
\mathbf{n}(x)=A\tr\bm{\tau}\Phi^{(1/2)}(x),
\end{equation}
where $A>0$ is a nonuniversal prefactor, and $\bm{\tau}$ is the vector of Pauli matrices. The spin-1/2 field has  scaling  dimension 3/8 and also enters into the staggered part of $\mathbf{S}_{j}\cdot\mathbf{S}_{j+1}$, from which we define the dimerization operator $\hat{d}\propto\tr\Phi^{(1/2)}$. We express the $\Phi^{(1/2)}$ operator as
\begin{equation}\label{eq:spin-1/2-operator}
\tr\Phi^{(1/2)}=\sigma^{1}\sigma^{2}\sigma^{3},\quad
\tr\tau^{a}\Phi^{(1/2)}=i\sigma^{a}\mu^{a+1}\mu^{a+2}.
\end{equation}
Here, $\sigma^{a}$ and $\mu^{a}$ are the Ising  order and disorder operators. We note that $\mu^{a}$ is fermionic in our notation. This means that it anticommutes with all fermion fields and other disorder operators \cite{Allen1997}. Another important observation is that our choice for the representation of $\Phi^{(1/2)}$ is not unique. As revealed by the duality transformation
\begin{equation}\label{eq:duality-transformation}
\xi^{a}\to\xi^{a},\qquad
\bar{\xi}^{a}\to-\bar{\xi}^{a},\qquad
\sigma^{a}\leftrightarrow\mu^{a},
\end{equation}
we could introduce a dual representation for $\Phi^{(1/2)}$ as well. The equivalence of this choice can be viewed as a gauge freedom, so that  adopting  the representation in Eq. (\ref{eq:spin-1/2-operator}) amounts to  fixing a gauge. We will come back to this point when discussing how to implement open boundary conditions.

The $\mathrm{SU(2)_{2}}$  WZW model has another scaling field. The spin-1 field $\Phi^{(1)}$ is a $3\times3$ matrix field with dimension 1. Its components are fermion bilinears of the form $\Phi^{(1)}_{ab}=i\xi^{a}\bar{\xi}^{b}$, so that its trace is given by the sum of energy operators:
\begin{equation}\label{eq:trace-spin-1-op}
\tr\Phi^{(1)}=\varepsilon^{1}+\varepsilon^{2}+\varepsilon^{3}.
\end{equation}
The $\mathrm{SU(2)}$-invariant trace of $\Phi^{(1)}$ appears in the smooth part of $\mathbf{S}_{j}\cdot\mathbf{S}_{j+1}$ and $(\mathbf{S}_{j}\cdot\mathbf{S}_{j+1})^{2}$. We can use this to learn about the vicinity of the $b=1$ critical point. For $|1-b|\ll1$, the spin-1 chain of Eq. (\ref{eq:spin-1-chain}) can be treated as a perturbed conformal field theory. The effective Hamiltonian includes the perturbation
\begin{equation}
\delta H=m\int dx\,\tr\Phi^{(1)},
\end{equation}
where $m\propto J(1-b)$ is a relevant coupling constant that governs the Haldane to dimerized transition. In the Ising model notation this transition is equivalent to a disorder to order transition in all Ising sectors. For   $m>0$,  the Ising models are disordered.  Although there are eight different   configurations for $\vev{\mu^{a}}\neq0$, the symmetry transformation
\begin{equation}
\xi^{a}\to-\xi^{a},\quad
\bar{\xi}^{a}\to-\bar{\xi}^{a},\quad
\sigma^{a}\to\sigma^{a},\quad
\mu^{a}\to-\mu^{a}
\end{equation} 
reduces this number to a physical fourfold degeneracy \cite{Allen2000}. This corresponds to the Haldane phase, characterized by the spontaneous breaking of the hidden $\mathbb{Z}_{2}\times\mathbb{Z}_{2}$ symmetry that gives a fourfold-degenerate ground state in an open chain \cite{Shelton1996}. In contrast, for   $m<0$ the Ising models are in the ordered phase. This case corresponds to the dimerized phase   with $\vev{\hat{d}\,}\propto\vev{\sigma^{1}\sigma^{2}\sigma^{3}}\neq0$.

We end this review by noting    that our construction based on the  $\mathrm{SU(2)_{2}}$  WZW model is not restricted to the specific lattice model of Eq. (\ref{eq:spin-1-chain}). The same critical theory applies, for instance,  to   the  spin-1 Heisenberg model with an additional three-site interaction:
\begin{align}\label{eq:spin-1-chain-3-site}
H_{\rm 3s}=&J\sum_{j}\Big\{(\mathbf{S}_{j}\cdot\mathbf{S}_{j+1})\nonumber\\
&+\beta\big[(\mathbf{S}_{j-1}\cdot\mathbf{S}_{j})(\mathbf{S}_{j}\cdot\mathbf{S}_{j+1})
+\hc\big]\Big\}.
\end{align}
This model is critical at $\beta\simeq0.111$   \cite{Michaud2012,Chepiga2016b}. The small critical ratio makes this a realistic model for the dimerization transition  in actual spin chain materials \cite{Michaud2012}.

%%%%%%%%%%%%%%%%%%%%%%%%%%%%%%%%%%%%%%%%%%%%%%%%%%%%%%%%%%

\subsection{Y junction}\label{secYjunction}

Let us now define the model of our junction. We consider a   Y junction that consists of three semi-infinite spin-1 chains coupled together only at their first sites. The Hamiltonian   is $H=H_{0}+H_{\mathcal{B}}$. The first term describes three critical spin-1 chains: 
\begin{equation}\label{eq:bulk-Hamiltonian}
H_{0}=J\sum_{j=1}^{\infty}\sum_{\alpha=1}^{3}\Big[(\mathbf{S}_{j,\alpha}\cdot\mathbf{S}_{j+1,\alpha})-(\mathbf{S}_{j,\alpha}\cdot\mathbf{S}_{j+1,\alpha})^{2}\Big],
\end{equation}
where $\mathbf{S}_{j,\alpha}$ is the spin-1 operator at site $j$ of chain $\alpha$. The second term describes the boundary interactions. We require it to preserve spin  SU(2) symmetry  and $\mathbb{Z}_{3}$ leg permutation symmetry, $\alpha\to\alpha+1$ (mod 3). We thus add to the model   a chiral interaction $J_{\chi}$ and an exchange interaction $J'$ between the end spins,
\begin{equation}\label{eq:boundary-Hamiltonian}
H_{\mathcal{B}}=J_{\chi}\hat{C}_{1}+J'\sum_{\alpha=1}^{3}\mathbf{S}_{1,\alpha}\cdot\mathbf{S}_{1,\alpha+1},
\end{equation}
where $\hat{C}_{j}=\mathbf{S}_{j,1}\cdot(\mathbf{S}_{j,2}\times\mathbf{S}_{j,3})$ is the scalar spin chirality operator at site $j$. The $J_{\chi}$ interaction breaks parity $\mathcal{P}:\alpha\to-\alpha$ and time reversal $\mathcal{T}: \mathbf{S}\to-\mathbf{S}$ symmetries, but preserves the combined $\mathcal{PT}$ symmetry. One may note that these symmetries also allow for a boundary biquadratic   interaction term $J_{q}'\sum_{\alpha}(\mathbf{S}_{1,\alpha}\cdot\mathbf{S}_{1,\alpha+1})^{2}$. However, we shall see that it suffices to tune two boundary parameters   to reach the chiral fixed point in the low-energy theory. For this reason, in this paper, we set $J_q'=0$  to reduce the parameter space of the model.

\begin{figure}
\centering
\includegraphics[width=.6\linewidth]{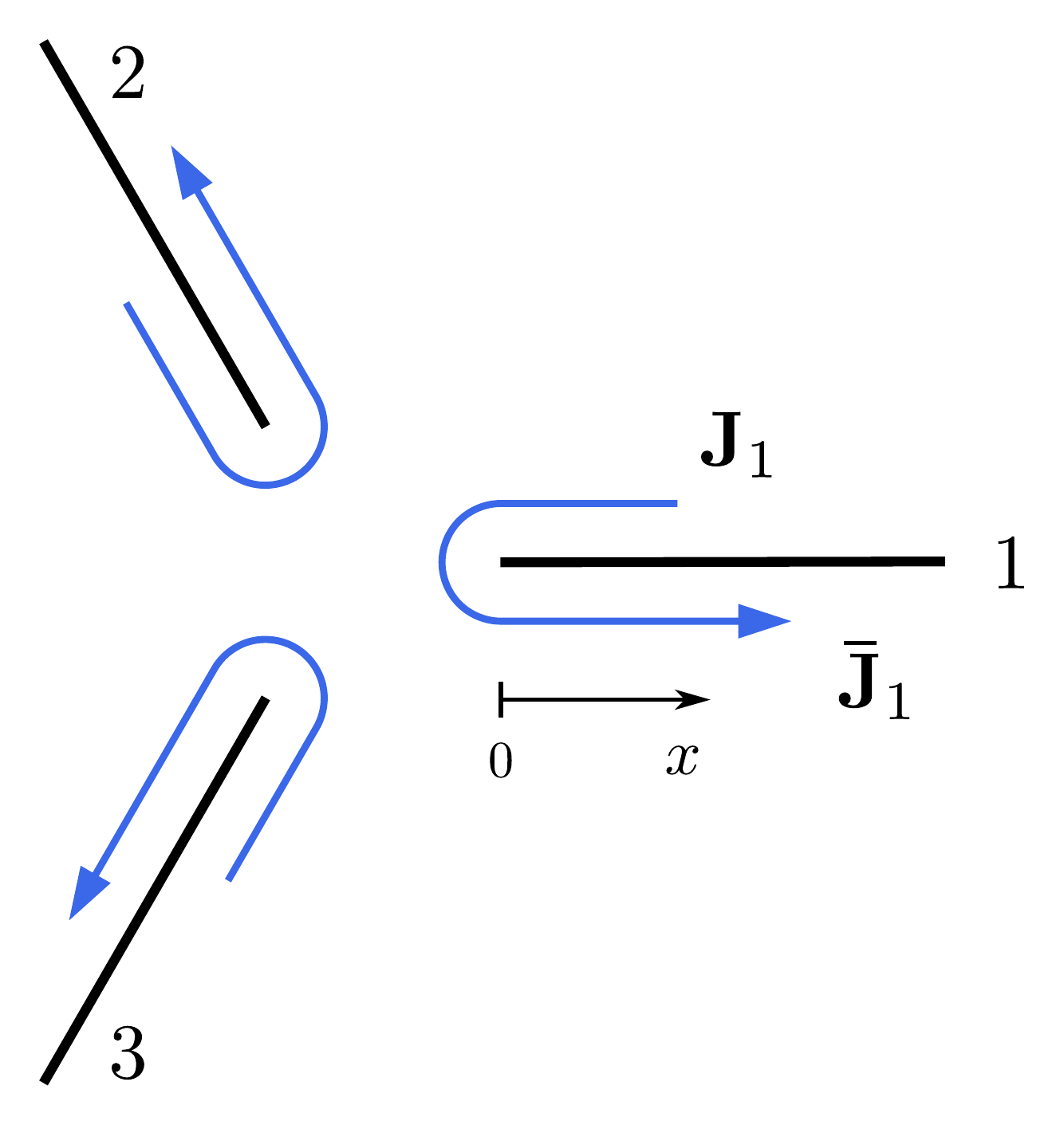}
\caption{Open-boundary fixed point.  The three legs are decoupled,   and incoming spin modes are fully reflected.}
\label{fig:open-junction} 
\end{figure}

We now consider the continuum limit of the Y junction. In the spirit of delayed evaluation of boundary conditions \cite{Oshikawa2006}, we write the bulk Hamiltonian   in Sugawara form as
\begin{equation}
H_{0}=\sum_{\alpha}\frac{\pi v}{2}\int_{0}^{\infty} dx \big( \mathbf{J}_{\alpha}^{2}+\mathbf{\bar{J}}_{\alpha}^{2}\big),
\end{equation}
where we dropped the marginal bulk interaction. In terms of the chiral Majorana fermions, we have \begin{equation}
H_0=\sum_{\alpha=1}^3\sum_{a=1}^3 \int_0^{\infty} dx\, \frac{iv}2(\xi_\alpha{}^{a}\nabla\xi_\alpha{}^{a}-\bar\xi_\alpha{}^{a}\nabla\bar\xi_\alpha{}^{a}),
\end{equation}
where $\nabla$ denotes the spatial derivative. We observe that, compared with the SU(2)$\times \mathbb{Z}_{3}$ symmetry of the   lattice model, this Hamiltonian has an enlarged SO(9)$\times$SO(9) chiral symmetry \cite{Witten1984}, which corresponds to rotations of the nine-component vectors $\bm{\xi}=(\xi_{1}{}^1,\xi_{1}{}^2,\dots,\xi_{3}{}^3)$ and $\bm{\bar{\xi}}=(\bar{\xi}_{1}{}^1,\bar\xi_{1}{}^2,\dots,\bar\xi_{3}{}^3)$. However,  once we  impose  boundary conditions  the chiral  currents  are no longer  independent. %Their relationship is going to be specified by the boundary conditions.
 By varying the boundary interactions we can drive  transitions between different  low-energy fixed points, which  must be identified with conformally invariant boundary conditions.  The simplest example   is the open ($\mathrm{O}$) fixed point    corresponding to three decoupled spin chains for   $J_\chi=J'=0$. The boundary conditions in this case are
\begin{equation}\label{eq:open-BC}
\mathbf{\bar{J}}_{\alpha}(x)=\mathbf{J}_{\alpha}(-x),
\end{equation}
meaning there is no flow across the boundary and incoming spin currents are fully reflected (see Fig. \ref{fig:open-junction}).

Let us now show how to deal with the boundary conditions in the Ising model formulation. From Eq. (\ref{eq:spin-currents}), we see that open boundary conditions can be implemented as either $\bar{\xi}_{\alpha}{}^{a}(x)=\xi_{\alpha}{}^{a}(-x)$ or $\bar{\xi}_{\alpha}{}^{a}(x)=-\xi_{\alpha}{}^{a}(-x)$. This sign ambiguity is a manifestation of the duality transformation in Eq. (\ref{eq:duality-transformation}), so that selecting a sign is equivalent to choosing a representation for $\Phi^{(1/2)}$. To be consistent with our choice in Eq. (\ref{eq:spin-1/2-operator}), we must use
\begin{equation}\label{eq:open-BC-Majorana}
\bar{\xi}_{\alpha}{}^{a}(x)=\xi_{\alpha}{}^{a}(-x).
\end{equation}
We derive the behavior of order and disorder operators by  combining two Ising models with the same leg index to define a bosonic field $\varphi_\alpha$ for each leg.  We employ the bosonization formulas 
\begin{equation}\label{eq:bosonization-formula}
\xi_{\alpha}{}^{1}+i\xi_{\alpha}{}^{2}=\frac{1}{\sqrt{\pi}}e^{i2\phi_{\alpha}},\qquad
\bar{\xi}_{\alpha}{}^{\,1}+i\bar{\xi}_{\alpha}{}^{\,2}=\frac{1}{\sqrt{\pi}}e^{i2\bar{\phi}_{\alpha}},
\end{equation}
where $\phi_{\alpha}$ and $\bar \phi_{\alpha}$ are the chiral components of the   boson $\varphi_{\alpha}=\phi_\alpha-\bar \phi_\alpha$; see Appendix \ref{app:bosonization} for details. From Eqs. (\ref{eq:open-BC-Majorana}) and  (\ref{eq:bosonization-formula}), we deduce that the open boundary conditions can be implemented by   
\begin{equation}\label{eq:open-BC-boson}
\bar{\phi}_{\alpha}(x)=\phi_{\alpha}(-x)+\mathcal{C},
\end{equation}
with $\mathcal{C}=0$ or $\mathcal{C}=\pi$. Substituting  this into the bosonized expressions
\begin{equation}\label{eq:bosonization-doubled-order}
\sigma_{\alpha}{}^{1}\sigma_{\alpha}{}^{2}=\cos\varphi_{\alpha},\qquad i\mu_{\alpha}{}^{1}\mu_{\alpha}{}^{2}=\sin\varphi_{\alpha},
\end{equation}
we find   $\vev{\sigma_{\alpha}{}^{a}}\propto x^{-1/8}$ and $\vev{\mu_{\alpha}{}^{a}}=0$. Thus we see that open boundary conditions are identified with fixed $\ket{\uparrow}$ or $\ket{\downarrow}$ boundary conditions in the Ising model representation \cite{Cardy1991}. Note that this leads to a nonvanishing expectation value for the dimerization, which decays away from the boundary as $\vev{\hat{d}_{\alpha}}\propto x^{-3/8}$. This behavior  is consistent with  the numerical study in Ref. \cite{Chepiga2016a}.

The next step is to analyze the stability of the O fixed point. If we impose open boundary conditions and express the local operators in terms of a single chiral component for each leg, Eq. (\ref{eq:spin-representation}) tells us that boundary spin operators are given by $\mathbf{S}_{1,\alpha}\propto\mathbf{J}_{\alpha}(0)$. We use this to write down  the leading boundary operators allowed by SU(2)$\times\mathbb Z_3$ symmetry that perturb the $\mathrm{O}$ fixed point:
\begin{align}\label{eq:boundary-perturbations-open}
H_{\mathcal{B}}^{\rm (O)}&=\kappa_{1}\sum_{\alpha}\mathbf{J}_{\alpha}(0)\cdot\mathbf{J}_{\alpha+1}(0)+\kappa_{2}\sum_{\alpha}[\mathbf{J}_{\alpha}(0)]^{2}\nonumber\\
&\quad+\kappa_{3}\sum_{\alpha}\mathbf{J}_{\alpha}(0)\cdot[\mathbf{J}_{\alpha+1}(0)\times\mathbf{J}_{\alpha-1}(0)]+\cdots,
\end{align}
where we obtain $\kappa_{1}\sim J'$ and $\kappa_{3}\sim J_{\chi}$ at weak coupling. Since each spin current $\mathbf{J}_{\alpha}$ has scaling dimension 1, all boundary perturbations in Eq. (\ref{eq:boundary-perturbations-open}) are irrelevant. We thus conclude that the $\mathrm{O}$ fixed point is stable and  governs the low-energy physics of the junction at weak coupling, $|J_\chi|, |J'|\ll J$.

To gain  some intuition about the other  fixed points that may appear in the  strong coupling limit  $|J_{\chi}|,|J'|\gg J$, we can   neglect the bulk Hamiltonian and diagonalize the three-spin Hamiltonian in Eq. (\ref{eq:boundary-Hamiltonian}).  Note that the eigenstates of $H_{\mathcal{B}}$  can be labeled by the eigenvalues of the total spin $\mathbf S_{\mathcal{B}}^2 =(\sum_\alpha \mathbf S_{1,\alpha})^2$, the $z$-spin component $S^z_{\mathcal B}$,  and the scalar spin chirality $\hat C_1$. Let us focus on $J'>0$. For $|J_\chi|/J'<\sqrt 3$, the ground state of $H_{\mathcal{B}}$ is a spin-singlet state with zero chirality and energy $E=-3J'$. Thus, in the limit $J\ll |J_\chi|\ll J'$, a simple picture for the junction consists of removing the three boundary spins  to form a singlet  and  coupling the spins at sites $j= 2$ by weak interactions generated by virtual transitions that excite the  singlet state. This picture suggests another fixed point of decoupled chains whose boundary has  been shifted by one site.

 For $|J_\chi|/J'>\sqrt 3$, the ground state of $H_{\mathcal{B}}$ becomes a  chiral spin-triplet state   with energy $E=-2J'-\sqrt{3}|J_{\chi}|$. The triplet ground state has negative  chirality if    $J_\chi>0$  and positive chirality if $J_\chi<0$. 
Let us denote by $\ket{\mathcal{B}_{m}}$ with $m\in\{+,0,-\}$ the three states in the ground state manifold. For $J_\chi>0$, the highest-weight state with maximum eigenvalue of $S^z_{\mathcal{B}}$ in this manifold is
\begin{align}
\ket{\mathcal{B}_{+}}&=\frac{1}{\sqrt{6}}\big(\ket{++-}+\omega\ket{-++}+\omega^{2}\ket{+-+}\nonumber\\
&\hspace{1.5cm}+\ket{00+}+\omega\ket{+00}+\omega^{2}\ket{0+0}\big),\label{triplet}
\end{align}
where $\omega=e^{i2\pi/3}$ and $\{\ket{m_{1},m_{2},m_{3}}\}$ is the basis of eigenstates  of $S^z_{1,\alpha}$. The other two states, $\ket{\mathcal{B}_{0}}$ and $\ket{\mathcal{B}_{-}}$, can be readily obtained from the action of the ladder  operator $S^{-}_{\mathcal B}$. Similarly, the time-reversal conjugates of the above states form  the ground state manifold of $H_{\mathcal B}$ for $J_\chi<0$. The picture for the junction in the limit $J\ll J'\ll |J_\chi|$   has an effective spin 1 at the boundary which must be symmetrically  coupled to the chains at the $j=2$ sites.  This limit is reminiscent of the Kondo   fixed point for the junction of spin-1/2 Heisenberg chains \cite{Buccheri2018}, where   the gapless bulk modes screen the emergent boundary spin  in close analogy with the three-channel Kondo problem.

%%%%%%%%%%%%%%%%%%%%%%%%%%%%%%%%%%%%%%%%%%%%%%%%%%%%%%%%%%

\section{Chiral fixed point}\label{sec:chiral-FP}

\begin{figure}
\centering
\includegraphics[width=.6\linewidth]{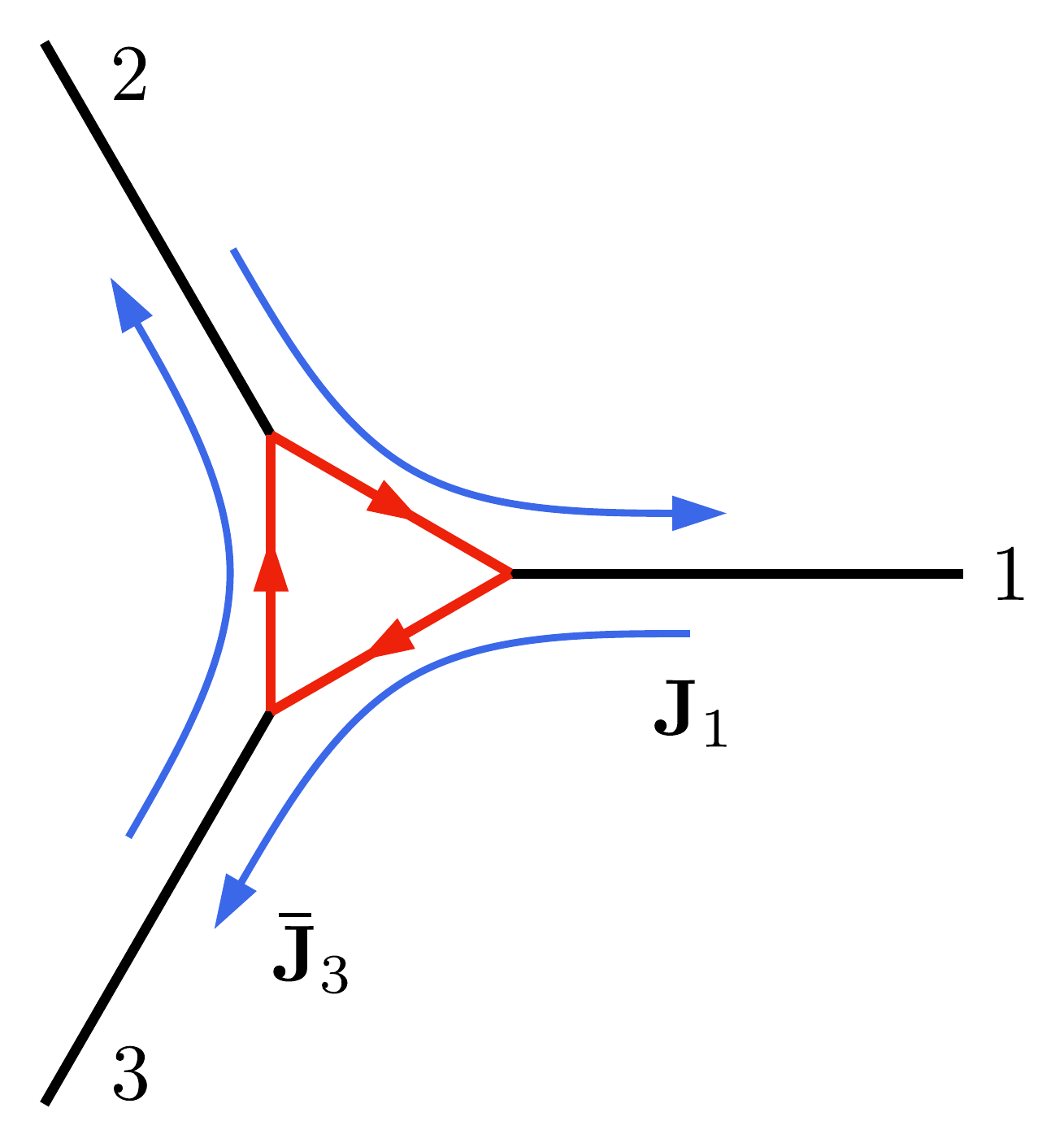}
\caption{Spin currents are diverted from chain $\alpha$ to chain $\alpha-1$ at the $\mathrm{C}_{-}$ fixed point.}
\label{fig:chiral-junction}
\end{figure}

When time reversal symmetry is broken, the junction may realize one of two chiral fixed points, $\mathrm{C}_{+}$ or $\mathrm{C}_{-}$. These are characterized by the complete transmission of incoming spin currents from one chain to the next in rotation. The circulation sense is determined by the sign of $J_{\chi}$. A positive $J_{\chi}$ favors a local negative-chirality state where spin currents circulate clockwise as shown in Fig. \ref{fig:chiral-junction}. This defines the $\mathrm{C}_{-}$ fixed point, described by the boundary conditions  
\begin{equation}\label{eq:chiral-BC}
\mathbf{\bar{J}}_{\alpha}(x)=\mathbf{J}_{\alpha+1}(-x).
\end{equation}
The circulation is reversed for negative $J_{\chi}$. Since both cases are clearly related by time reversal, without loss of generality, we  consider $J_{\chi}>0$ from now on.

At the chiral fixed point, the spin chirality $\hat{C}_{j}$ acquires  a nonzero expectation value \cite{Buccheri2018}. To see how this follows from the boundary conditions in Eq. (\ref{eq:chiral-BC}), we calculate the large-distance decay of $\vev{\hat{C}_{j}}$. Since in the continuum limit the most relevant contribution to $\hat{C}_{j}$ stems from the staggered magnetization fields, the evaluation of this correlator boils down to considering $\vev{\hat{C}_{j}}\sim(-1)^{j}\vev{\mathbf{n}_{1}\cdot(\mathbf{n}_{2}\times\mathbf{n}_{3})}$. We   proceed with  the Ising model representation of the staggered components in Eqs. (\ref{stagg}) and (\ref{eq:spin-1/2-operator}). Let us then introduce the $\chi$ operator $\chi_{123}{}^{a}=i\mu_{1}{}^{a}\mu_{2}{}^{a}\sigma_{3}{}^{a}$, so that the contribution from the staggered magnetization fields to $\hat{C}_{j}$ is written as
\begin{align}\label{eq:spin-chirality-chi-rep}
\hat{C}_{j}\sim A^{3}(-1)^{j+1}\big(&\chi_{231}{}^{1}\chi_{312}{}^{2}+\chi_{312}{}^{1}\chi_{231}{}^{2}\big)\chi_{123}{}^{3}\nonumber\\
&+\cyclic+\cdots.
\end{align}
Next, we use bosonization to compute its expectation value. As for the $\mathrm{O}$ fixed point, we have two ways to implement chiral boundary conditions in the Ising model formulation, $\bar{\xi}_{\alpha}{}^{a}(x)=\pm\xi_{\alpha+1}{}^{a}(-x)$. We stress, once again, that both ways are equivalent and reflect the gauge transformation in Eq. (\ref{eq:duality-transformation}). However, as we are working in the fixed gauge of Eq. (\ref{eq:spin-1/2-operator}), we adopt
\begin{equation}\label{eq:chiral-BC-Majorana}
\bar{\xi}_{\alpha}{}^{a}(x)=\xi_{\alpha+1}{}^{a}(-x).
\end{equation}
In the bosonic representation, the above boundary conditions  are equivalent to imposing 
\begin{equation}\label{eq:chiral-BC-boson}
\bar{\phi}_{\alpha}(x)=\phi_{\alpha+1}(-x)+\mathcal{C},
\end{equation}
with $\mathcal{C}=0$ or $\mathcal{C}=\pi$. We are now in position to compute the expectation value of a pair of $\chi$ operators, which according to Eq. (\ref{eq:bosonization-doubled-order}) takes the form
\begin{equation}
\chi_{123}{}^{1}\chi_{123}{}^{2}=-\sin\varphi_{1}\sin\varphi_{2}\cos\varphi_{3}.
\end{equation}
Plugging the boundary conditions of Eq. (\ref{eq:chiral-BC-boson}) into this formula, we obtain $\vev{\chi_{123}{}^{a}}\propto x^{-3/8}$. If we now return to Eq. (\ref{eq:spin-chirality-chi-rep}), we conclude that the spin chirality has the power law decay
\begin{equation}
\vev{\hat{C}_{j}}\propto(-1)^{j}x^{-9/8}.
\end{equation}
Note that the long-distance decay of this correlation function is governed by the bulk scaling dimension of $\hat{C}_{j}$, as expected for the chiral fixed point. We remark that, although we may observe a nonzero $\vev{\hat{C}_{j}}$ whenever we have  $J_{\chi}\neq0$ in the lattice model, the key feature is that the chiral fixed point  has the slowest decay for $\vev{\hat{C}_{j}}$. In the regimes governed by nonchiral fixed points, the nonzero expectation value of  $\vev{\hat{C}_{j}}$ depends on irrelevant boundary operators. For instance,  near the $\mathrm{O}$ fixed point, the leading chiral boundary operator has dimension 3, and standard perturbation theory to the first order in $\kappa_3$ in Eq. (\ref{eq:boundary-perturbations-open}) yields $\vev{\hat{C}_{j}}\propto(-1)^{j}x^{-25/8}$ instead.

We now turn to the stability of the chiral fixed point. The boundary perturbations are determined by the operator content of the conformal  field theory. The leading boundary operators   respecting SU(2)$\times \mathbb Z_3$ symmetry are constructed from the trace of the primary matrix fields. For the junction of $\mathrm{SU}(2)_{2}$ WZW models,  the  boundary operators that we must pay attention  to  are 
\begin{equation}
H_{\mathcal{B}}^{\rm (C)}=\lambda_{1}\sum_{\alpha=1}^{3}\tr\Phi^{(1/2)}_{\alpha}(0)+\lambda_{2}\sum_{\alpha=1}^{3}\tr\Phi^{(1)}_{\alpha}(0).
\end{equation}
The first term is a relevant perturbation of dimension 3/8 and the second one is a dimension-1 marginal perturbation. Here we neglect irrelevant boundary operators written in terms of the chiral currents.

We can understand how the leading  boundary operators are generated in our model  by approaching the problem from   weak coupling. We first look at the chiral boundary interaction, $H_{\chi}=J_{\chi}\hat{C}_{1}$.  We separate the contributions with different factors of uniform or staggered components of the spin operators contained in $\hat C_1$. The contribution with three factors of the staggered magnetization  generates the marginal perturbation
\begin{equation}
H^{ (nnn)}_{\chi}\sim J_{\chi}\frac{A^{3}}{4}\left[-3+\pi\sum_{\alpha=1}^{3}\tr\Phi^{(1)}_{\alpha}(0)\right]+\cdots.
\end{equation}
Its leading term is proportional to the identity, which is consistent with the fact that $H_{\chi}$ should favor the formation of a chiral boundary state. To obtain this result, we expanded the $\chi$ operators in Eq. (\ref{eq:spin-chirality-chi-rep}) around $x=0$ and then used bosonization to show that
\begin{equation}\label{eq:chi-OPE}
\chi_{123}(0)\sim\pm\frac{1}{2}\Big\{1-\pi\big[\varepsilon_{1}(0)+\varepsilon_{2}(0)-\varepsilon_{3}(0)\big]\Big\}+\cdots.
\end{equation}
Here we have omitted the spin index for brevity. Note that, due to the duplication of the model, there is an overall sign which is not fixed by bosonization. In face of this, we choose the sign so that the total energy is lowered, $\vev{H_{\chi}}<0$. The chiral boundary interaction also generates the relevant operator from the contribution that mixes the staggered magnetization with chiral currents,
\begin{align}\label{eq:chirality-JJn}
H^{(JJn)}_{\chi}&\sim-J_{\chi}\left[J^{x}_{1}(0)\bar{J}^{y}_{2}(0)-J^{y}_{1}(0)\bar{J}^{x}_{2}(0)\right]n^{z}_{3}(0)\nonumber\\
&\qquad+\cyclic.
\end{align}
Imposing  the chiral boundary conditions in Eq. (\ref{eq:chiral-BC}), we can rewrite this term as a product of operators with the same leg index:
\begin{align}\label{eq:chirality-mixed-after-BC}
H^{(JJn)}_{\chi}&\sim-J_{\chi}\big[\bar{J}^{x}_{3}(0){J}^{y}_{3}(0)-\bar{J}^{y}_{3}(0){J}^{x}_{3}(0)\big]n^{z}_{3}(0)\nonumber\\
&\qquad+\cyclic.
\end{align}
This trick is helpful as long as $\mathbf{J}_{\alpha}$ and $\mathbf{\bar{J}}_{\alpha}$ are independent for a given $\alpha$, so we can use the standard operator product expansions (OPEs) for a single spin chain. We obtain the relevant perturbation
\begin{equation}
H^{(JJn)}_{\chi}\sim J_{\chi}\frac{A}{8\pi^{2}}\sum_{\alpha=1}^{3}\tr\Phi^{(1/2)}_{\alpha}(0)+\cdots.
\end{equation}
As a result, the chiral three-spin interaction produces both relevant and marginal boundary operators. Next, we consider the exchange boundary interaction $H_{1}=J'\sum_{\alpha}\mathbf{S}_{\alpha}\cdot\mathbf{S}_{\alpha+1}$. The latter only produces the relevant boundary operator at the first order. From the fusion of the staggered magnetization fields, we obtain
\begin{equation}
H^{(nn)}_{1}\sim J'\frac{3A^{2}}{\sqrt{8}}\sum_{\alpha=1}^{3}\tr\Phi^{(1/2)}_{\alpha}(0)+\cdots.
\end{equation}
We thus write symbolically $\lambda_{1}\sim J_{\chi}+J'$ and $\lambda_{2}\sim J_{\chi}$.

This result indicates that the chiral fixed point cannot appear at weak coupling because  it is destabilized by a relevant perturbation. However,   it can still  appear as an intermediate-coupling fixed point associated with a boundary phase transition in our  model.   %by analyzing the effects of the leading perturbations in the effective field theory. 
 To reach the chiral fixed point, we need to tune two parameters in the lattice model so that both $\lambda_1$ and $\lambda_2$ vanish. 
%In face of this, let us use our physical intuition to look for the chiral fixed point at intermediate coupling. Suppose we fine tune one parameter in the lattice model so that the relevant coupling constant $\lambda_{1}$ vanishes. Since we still would have a marginal perturbation left, we will need to tune another parameter to suppress $\lambda_{2}$ as well. 
 Let us   assume that $\lambda_1$ and $\lambda_2$ are smooth functions of the boundary couplings $J_\chi$ and $J'$ and  that  the chiral fixed point is realized at some critical values  $(J_{\chi,c},J'_{c})$. Near this putative point, the effective couplings behave as
\begin{align}
\lambda_{1}&\approx -c_{1}(J_{\chi}-J_{\chi,c})-c_{2}(J'-J'_{c}),\nonumber\\
\lambda_{2}&\approx -c_{3}(J_{\chi}-J_{\chi,c}).
\end{align}
Here $c_{i}$ are nonuniversal positive constants so that $\lambda_1$ and $\lambda_2$ are positive at weak coupling, in agreement with the perturbative expressions. To connect the effective field theory with chiral boundary conditions to the picture for the weak and strong coupling limits discussed in Sec. \ref{secYjunction}, we next address what happens when $\lambda_1$ and $\lambda_2$ change sign across the chiral fixed point.

%%%%%%%%%%%%%%%%%%%%%%%%%%%%%%%%%%%%%%%%%%%%%%%%%%%%%%%%%%

\section{Boundary  phase diagram}\label{sec:relevant}

In this section we analyze the effects of the relevant and marginal perturbations to the chiral fixed point. Our aim is to substantiate the   phase diagram sketched   in Fig. \ref{fig:boundary-phase-diagram}, which places the chiral fixed point   on a transition line between two phases governed by open-boundary fixed points.  

%%%%%%%%%%%%%%%%%%%%%%%%%%%%%%%%%%%%%%%%%%%%%%%%%%%%%%%%%%

\subsection{Relevant operator}

We first examine the   relevant boundary operator. Under the RG the effective coupling $\lambda_{1}$ flows monotonically toward strong coupling, dominating the infrared behavior of the junction. When this happens, we must abandon the chiral boundary conditions and pin the boundary fields to minimize the  term $\lambda_{1}\sum_{\alpha}\tr\Phi^{(1/2)}_{\alpha}(0)$. Since the trace of the spin-1/2 matrix field defines the dimerization $\hat{d}\propto\tr\Phi^{(1/2)}$, we see that  $\lambda_{1}$ leads to a nonzero expectation value for the boundary dimerization $\vev{\hat{d}_{\alpha}(0)}\neq0$. This behavior is representative of the O fixed point discussed   in Sec. \ref{secYjunction}. 

\begin{figure}
\centering
\includegraphics[width=1\linewidth]{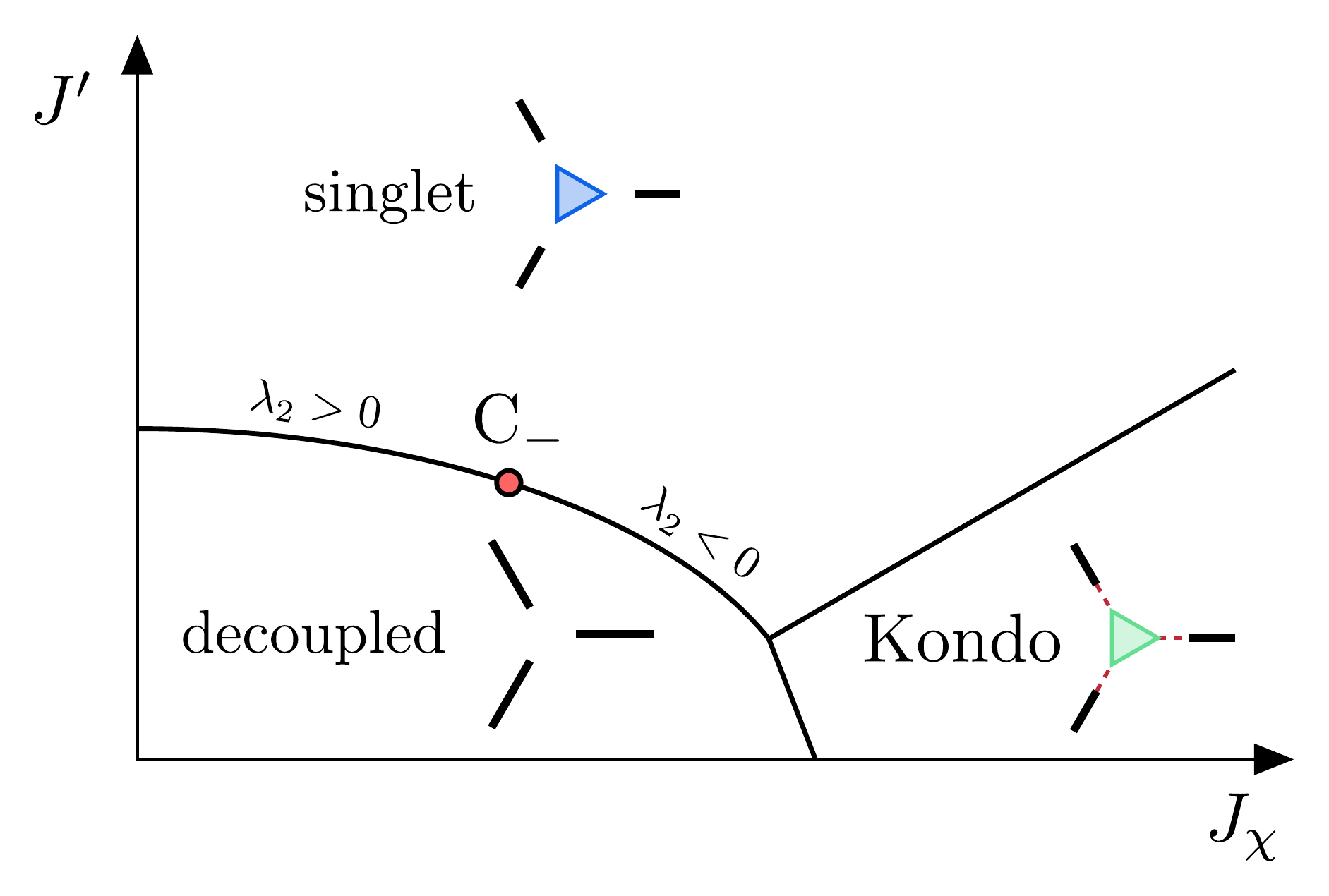}
\caption{Schematic boundary phase diagram for the junction. The chiral fixed point is denoted by a red dot and lies on the transition line that separates the decoupled and singlet phases. In addition to these, there is a Kondo-like phase   in the large-$J_{\chi}$ limit.}
\label{fig:boundary-phase-diagram}
\end{figure}

%Thus $\lambda_{1}$ governs the transition between two phases described by open boundary conditions. To see that, 

To understand what happens when $\lambda_1$ changes sign, we recall that the expectation value of the dimerization controls the relative strength of  even and odd bonds in the spin chains:
\begin{equation}
\vev{\mathbf{S}_{j,\alpha}\cdot\mathbf{S}_{j+1,\alpha}}=C_{\rm unf}(x)+(-1)^{j}\vev{\hat{d}_{\alpha}(x)},
\end{equation}
where $C_{\rm unf}(x)<0$ denotes the uniform part of the spin-spin correlation, which is a smooth function of $x$.  The  sign of $\vev{\hat{d}_{\alpha}(x)}$ depends on the sign of $\lambda_1$, which  flows to either $\lambda_1\to \infty $ or $\lambda_1\to-\infty$.  At weak coupling, we have $\lambda_1>0$. This   corresponds to the O fixed point of trivially decoupled chains, at which the first bond of each chain  is stronger than the second bond. In contrast, for $\lambda_1<0$ the second bond becomes the strongest.  This case is compatible with  the  regime of dominant   $J'>0$ discussed   in Sec. \ref{secYjunction}. The sign change of $\lambda_{1}$ can then  be associated with the formation of a boundary spin singlet, which is weakly coupled to the rest of the chains in the limit $J'\gg J_\chi,J$.  This picture bears resemblance to the two-impurity Kondo model in the presence of particle-hole symmetry \cite{AffleckJones1995,Zarand2006}. In the latter, a quantum phase transition separates two disconnected phases   corresponding to the regime where each impurity is screened by  its neighboring chain and the regime where the pair of impurities forms a singlet state. We note that the two-impurity Kondo model in spin chains was studied in Refs. \cite{Bayat2012,Bayat2014}. Based on these observations, we claim that the relevant coupling $\lambda_{1}$ drives a transition from the decoupled to the singlet phase of our junction, see Fig. \ref{fig:boundary-phase-diagram}.

%To construct the phase diagram of our junction, we need to identify candidates for the infrared fixed points. Since the exact diagonalization of $H_{\mathcal{B}}$ yields a spin-triplet for $|J_{\chi}|\to\infty$ and $J'>0$, a simple guess is that we shall find a Kondo-like fixed point in the low-energy limit. With this in mind, we sketch the phase diagram in Fig. \ref{fig:boundary-phase-diagram}. In the following we enrich our view about the phase diagram by studying the marginal boundary interaction, which reshuffles spin modes across the junction and leads to very interesting physics.

%%%%%%%%%%%%%%%%%%%%%%%%%%%%%%%%%%%%%%%%%%%%%%%%%%%%%%%%%%

\subsection{Marginal operator and spin conductance }\label{sec:marginal}

We now move away from the chiral fixed point along the phase transition line, so that the only perturbation  is the marginal boundary interaction. The effective Hamiltonian with $\lambda_1=0$, $\lambda_2\neq 0$ takes a rewardingly  simple form in  the Ising model representation. Since the trace of $\Phi^{(1)}$ in Eq. (\ref{eq:trace-spin-1-op}) does not mix different spin components, the Majorana fermions   with different  spin  indices are all independent, and we can treat  them separately. The effective Hamiltonian for a single spin component reads  
\begin{equation}
H_c=\sum_{\alpha}\int_{-\infty}^{\infty} dx\bigg[\frac{iv}{2}\xi_{\alpha}\nabla\xi_{\alpha}+i\lambda_{2}\delta(x)\xi_{\alpha}\xi_{\alpha+1}\bigg],
\end{equation}
where we have used the chiral boundary conditions to unfold the space interval and omitted the spin index in the Majorana fermions to lighten the notation. This Hamiltonian describes three left-moving Majorana fermions with a pointlike scattering amplitude at $x=0$. Since the  Hamiltonian is quadratic, we can diagonalize it  exactly and use the result to calculate  the spin transport properties  of the junction along the transition line.

%%%%%%%%%%%%%%%%%%%%%%%%%%%%%%%%%%%%%%%%%%%%%%%%%%%%%%%%%%

To solve the scattering  problem, it is   convenient to change basis. Using the three Majorana fermions, we construct one real fermion $\xi_{0}$ and one complex fermion $\psi$. The real fermion is defined as
\begin{equation}
\xi_{0}=\frac{1}{\sqrt{3}}(\xi_{1}+\xi_{2}+\xi_{3}).
\end{equation}
The complex fermion and its conjugate are written as
\begin{align}
\psi=\frac{1}{\sqrt{3}}(\xi_{1}+\omega\xi_{2}+\omega^{2}\xi_{3}),\nonumber\\
\psi^{\dagger}=\frac{1}{\sqrt{3}}(\xi_{1}+\omega^{2}\xi_{2}+\omega\xi_{3}).
\end{align}
Note that $\xi_0$ is invariant, while $\psi^\dagger$ transforms as $\psi^\dagger \to\omega \psi^\dagger$ under   cyclic leg permutations $\alpha\to\alpha+1$. In terms of the new fermions, the Hamiltonian is decomposed into two independent terms:
\begin{equation}\label{eq:chiral-plus-marginal}
H_c=\frac{i}{2}\int dx\,\xi_{0}\nabla\xi_{0}+\int dx\,\psi^{\dagger}[i\nabla+\gamma\delta(x)]\psi,
\end{equation}
with $\gamma=\sqrt{3}\lambda_{2}$ and  $v$ set to unity. The first term describes  a free, left-moving Majorana fermion, which evolves in real time as
\begin{equation}
\xi_{0}(x,t)=\int\frac{dk}{2\pi}\,b_{k}e^{-ik(t+x)},
\end{equation}
where $k$ is the momentum. The reality condition on $\xi_{0}$ implies that  the normal-mode operators satisfy $b^{\phantom\dagger}_{-k}=b_{k}^{\dagger}$ with $\acomm{b_{k}^{\phantom\dagger}}{b_{k'}^{\dagger}}=2\pi\delta(k-k')$. The second term in Eq.  (\ref{eq:chiral-plus-marginal}) describes the potential scattering  of the complex fermion at the origin. The equation of motion  is
\begin{equation}\label{eq:EOM-psi-impurity}
i(\partial_{t}-\nabla)\psi=\gamma\delta(x)\psi.
\end{equation}
Since $\psi$ satisfies a free equation of motion for $x\neq0$, we can write the solution in terms of  two plane waves, one for $x<0$ and the other for $x>0$. To match the solutions on both sides, we integrate the equation of motion over the infinitesimal interval $x\in(-\epsilon,+\epsilon)$ and obtain
\begin{equation}
\psi(0^{+})=\frac{1+i\frac{\gamma}{2}}{1-i\frac{\gamma}{2}}\psi(0^{-}).
\end{equation}
As a result, the mode expansion for the complex fermion  takes the form
\begin{equation}
\psi(x,t)=e^{2i\delta \Theta(x)}\int \frac{dk}{2\pi}\, c_{k}e^{-ik(t+x)}
\end{equation}
with $\acomm{c_{k}^{\phantom\dagger}}{c_{k'}^{\dagger}}=2\pi\delta(k-k')$ and $\Theta(x)$  the Heaviside step function. The phase shift $\delta$ is set by the condition $\tan\delta=\gamma/2$, saturating at $\delta\to\pm\pi/2$ as we send the coupling to $\gamma\to\pm\infty$. Furthermore, for small $\gamma$, we can approximate $\delta\approx\gamma/2$. Importantly, the phase shift is a smooth function of $\gamma$, and there is no  discontinuity across  the chiral fixed point where $\gamma$ changes sign.

The seemingly innocuous phase shift controls the transmission rate of Majorana modes across the junction. To see that, we return to the original Majorana basis and consider two-point functions of the form $\vev{\bar{\xi}_{\alpha}(\bar{z})\xi_{\beta}(w)}$, where  $z$ and $\bar{z}$ are complex coordinates in Euclidean spacetime, defined as $z=\tau+ix$ and $\bar{z}=\tau-ix$ for imaginary time $\tau$. Under the chiral boundary conditions of Eq. (\ref{eq:chiral-BC-Majorana}), these correlators correspond to $\vev{\xi_{\alpha+1}\xi_{\beta}}$ with operators taken at opposite sides of the origin. Using the mode expansions for $\xi_{0}$ and $\psi$, we   obtain
\begin{equation}\label{eq:right-left-majorana-correlator}
\vev{\bar{\xi}_{\alpha}(\bar{z})\xi_{\beta}(w)}=\frac{t_{\alpha\beta}}{2\pi(\bar{z}-w)},
\end{equation}
where the coefficients $t_{\alpha\beta}$ are functions of the phase shift:
\begin{align}
t_{11}&=\frac{1+2\cos(2\delta+\tfrac{2\pi}{3})}{3},\nonumber\\
t_{21}&=\frac{1+2\cos(2\delta-\tfrac{2\pi}{3})}{3},\nonumber\\
t_{31}&=\frac{1+2\cos(2\delta)}{3}.
\end{align}
The other components of $t_{\alpha\beta}$ can be obtained by cyclic permutation of the indices.  Note that for $\delta=0$ we correctly recover   $t_{31}=1$ and $t_{11}=t_{21}=0$, as expected  for the $\mathrm{C}_{-}$ fixed point.

We are now ready to address the spin transport of the chiral fixed point deformed by the marginal interaction. The central quantity here is the spin conductance tensor \cite{Buccheri2019}, which can be probed by measuring the response of the junction to the application of a spin chemical potential $\mathbf{B}_{\alpha}$ at the end of  chain $\alpha$. Within linear response theory, we   have
\begin{equation}
\vev{I_{\alpha}^{a}}=\sum_{b,\beta}G_{\alpha\beta}^{ab}B_{\beta}^{b},
\end{equation}
where $\mathbf{I}_{\alpha}(x)=\mathbf{J}_{\alpha}(x)-\mathbf{\bar{J}}_{\alpha}(x)$ is the spin current flowing into the junction from chain $\alpha$, and $G_{\alpha\beta}^{ab}$ are the components of the spin conductance tensor. The spin conductance can be obtained from the Kubo formula \cite{Oshikawa2006}
\begin{equation}\label{eq:Kubo-formula}
G_{\alpha\beta}^{ab}=-\lim_{\omega\to0^{+}}\frac{1}{\omega L}\int_{0}^{L}dx\int_{-\infty}^{\infty}d\tau\, e^{i\omega\tau}\vev{I_{\alpha}^{a}(x,\tau)I_{\beta}^{b}(y,0)},
\end{equation}
with arbitrary $y>0$. For instance, let us compute the spin conductance between two distinct chains $\alpha\neq\beta$. Our calculations follow closely those of Ref. \cite{Rahmani2012}. The nonzero contribution to the two-point function $\vev{I_{\alpha}^{a}I_{\beta}^{b}}$ comes from the left-right correlations
\begin{align}
\vev{I_{\alpha}^{a}(x,\tau)I_{\beta}^{b}(y,0)}=-\big[&\vev{J_{\alpha}^{a}(x,\tau)\bar{J}_{\beta}^{b}(y,0)}\nonumber\\
&+\vev{\bar{J}_{\alpha}^{a}(x,\tau)J_{\beta}^{b}(y,0)}\big].
\end{align}
Using Eq. (\ref{eq:right-left-majorana-correlator}), we find that these two-point correlators are given by
\begin{equation}
\vev{I_{\alpha}^{a}(x,\tau)I_{\beta}^{b}(y,0)}=-\frac{\delta^{ab}}{4\pi^{2}}\bigg[\frac{t_{\beta\alpha}^{2}}{(z-\bar{w})^{2}}+\frac{t_{\alpha\beta}^{2}}{(\bar{z}-w)^{2}}\bigg],
\end{equation}
where $z=\tau+ix$ and $w=iy$. Note that  SU(2) symmetry implies that the conductance is diagonal in the spin indices, and we write $G_{\alpha\beta}^{ab}=\delta^{ab}G_{\alpha\beta}$.  Plugging the result for the  correlator back into Eq. (\ref{eq:Kubo-formula}) and performing the   integrals, we obtain  
\begin{equation}
G_{\alpha\beta}=-\frac{1}{2\pi}t_{\alpha\beta}^{2},\qquad \alpha\neq\beta.
\end{equation}
The calculation for $\alpha=\beta$ is quite similar and yields
\begin{equation}
G_{\alpha\alpha}=\frac{1}{2\pi}(1-t_{\alpha\alpha}^{2}).
\end{equation}
Note that $G_0=1/(2\pi)$ is the quantum of spin conductance in units of $\hbar=1$ \cite{Yao2011}. One simple check here is to verify that our formulas   satisfy $\sum_{\alpha}G_{\alpha\beta}=\sum_{\beta}G_{\alpha\beta}=0$, as required from spin current conservation $\sum_{\alpha}\vev{\mathbf{I}_{\alpha}}=0$ and the fact that applying the same spin chemical potential  $\mathbf{B}_{\alpha}$ to all chains leads to zero current.

\begin{figure}
\centering
\includegraphics[width=.9\linewidth]{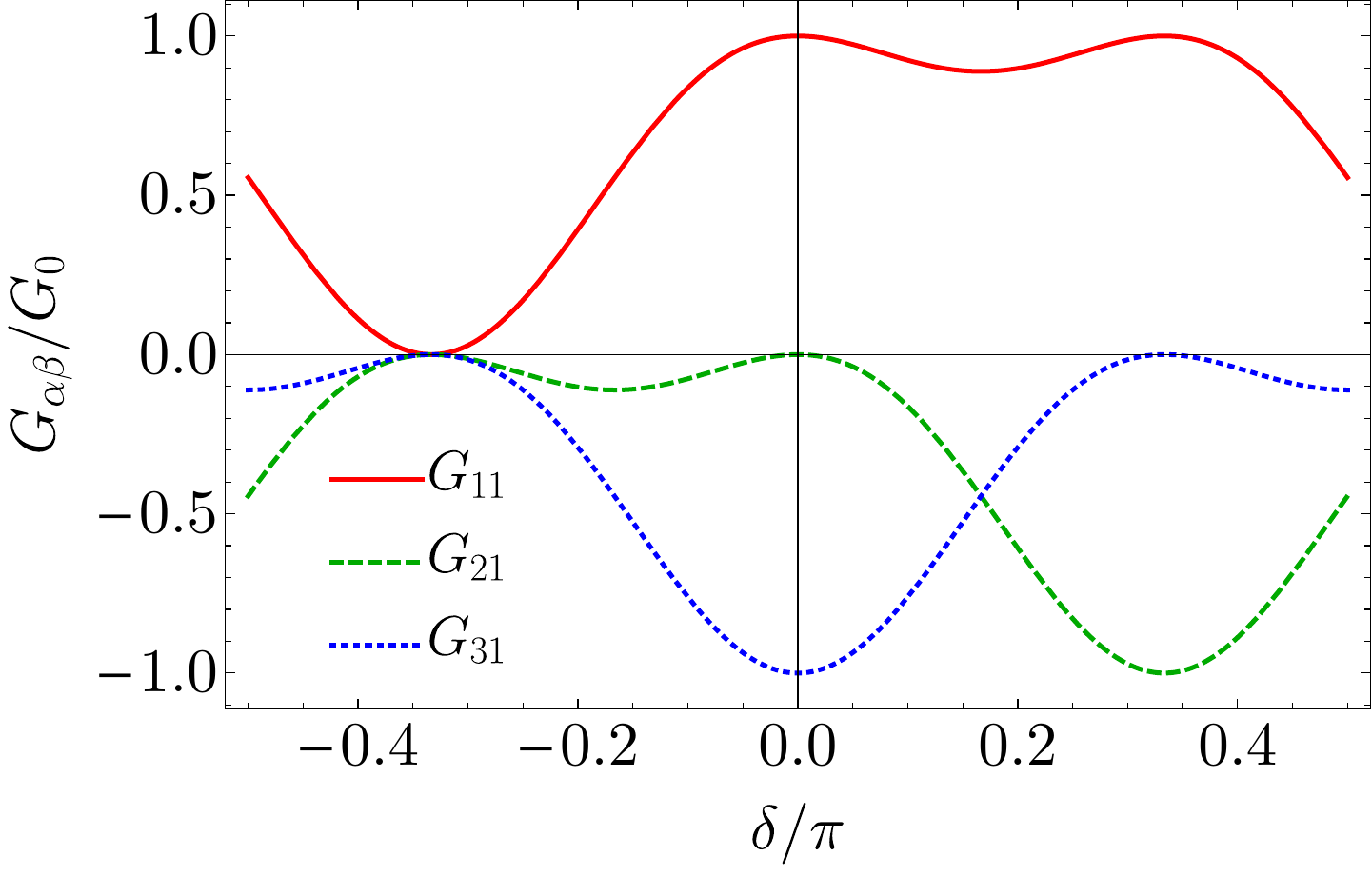} 
\caption{Components of the spin conductance tensor, in units of     $G_{0}=1/2\pi$, as a function of the phase shift $\delta$.}
\label{fig:spin-conductance}
\end{figure}

The spin conductance as a function of the phase shift is shown in Fig. \ref{fig:spin-conductance}. At  $\delta=0$, we have $G_{11}=-G_{31}=G_0$ and $G_{21}=0$. Defining the conductance asymmetry as $G_{\rm A}=(G_{31}-G_{21})/G_0$, we find  $G_{\rm A}=-1$ for  $\delta =0$, so  that the spin conductance is maximally asymmetric at the chiral  fixed point.  While the conductance varies continuously with the phase shift,   positive and negative values of $\delta$ lead to rather distinct behaviors. Let us explore this difference to uncover the role of the marginal coupling $\lambda_{2}$ on the phase diagram of Fig. \ref{fig:boundary-phase-diagram}. Departing from $\delta=0$ toward positive phase shifts, we encounter two points that draw our attention. The first one occurs at $\delta=\pi/6$,  where an incoming  spin current  is transmitted symmetrically to the other two chains,   $G_{21}=G_{31}$, and parity symmetry  is restored. If we increase $\delta$ further, the junction starts to work with a reversed chirality, so that $|G_{21}|>|G_{31}|$.  We  encounter the second interesting point   at $\delta=\pi/3$, where the junction is once again an ideal spin circulator, but now $G_{\rm A}=+1$, and  the currents  are rerouted in the counterclockwise direction. 

Since the perturbative expressions derived in Sec. \ref{sec:chiral-FP} indicated that $\lambda_2>0$ at weak coupling, we infer that moving along the transition line in the direction of positive $\delta$ must correspond to decreasing $J_\chi$ (see Fig. \ref{fig:boundary-phase-diagram}). In particular, it is tempting to identify   $\delta=\pi/6$ with the   point where the transition line crosses the $J'$ axis and  $\mathcal P$ and $\mathcal T$ symmetries are restored because $J_\chi=0$. According to the relation between the phase shift and the scattering amplitude in the effective model, for $\delta=\pi/6$, we have $\lambda_2=2/3$, a sizable  deviation from the C$_-$   fixed point.  Remarkably,   $G_{\alpha\alpha}/G_0=8/9$ for $\delta=\pi/6$ is the maximum  value of the conductance for free fermions in a three-lead junction allowed by unitarity and time-reversal invariance \cite{Nayak1999}. We conjecture that this nontrivial  value describes the spin conductance  of our  junction  at the transition between the two open-boundary fixed points for  $J_\chi=0$. It is also interesting to note that the result  for the conductance  in Fig.  \ref{fig:spin-conductance} is symmetric about $\delta=\pi/6$ if we also exchange $G_{21}$ and $G_{31}$. Thus, if we extrapolate the result beyond the parity-symmetric point, it makes sense to associate $\delta=\pi/3$ with the C$_+$   fixed point found  on the $J_\chi<0$ side of the phase diagram.

%%%%%%%%%%%%%%%%%%%%%%%%%%%%%%%%%%%%%%%%%%%%%%%%%%%%%%%%%%  

\subsection{Boundary bound state}\label{sec:bound}

Let us now consider negative values of the phase shift. Note that, according to Eq. (\ref{eq:chiral-plus-marginal}), $\lambda_2<0$ is related to an attractive   potential for the $\mathbb Z_3$-charged complex fermion. As we move away from $\delta=0$, the spin conductance is rapidly suppressed along this direction, culminating at $\delta=-\pi/3$ where the junction becomes disconnected with all $G_{\alpha\beta}=0$. However,   this point takes place at $\lambda_{2}=-2$, far from the perturbative regime.  Here, we should take into account the possibility that such a strongly attractive potential may lead to the formation of a bound state whose wave function is localized near $x=0$. This effect is not captured by the   effective Hamiltonian of Eq. (\ref{eq:chiral-plus-marginal}). Unlike the textbook example of a particle with quadratic dispersion described by the Schr\"odinger equation, a chiral fermion with linear dispersion does not develop a bound state for an arbitrarily weak  potential well in one dimension. In the many-body problem, bound states can be detected  as poles in the two-particle propagator \cite{mahan1990many}, but we have verified that there are no such poles in  the case of our effective Hamiltonian. However, this conclusion may break down when the attractive potential becomes comparable with the high-energy cutoff  in the effective field theory so that the linear dispersion approximation is no longer applicable. In Appendix \ref{app:bound-state} we show that  two different lattice regularizations of the fermionic model which reduce to the second term in  Eq. (\ref{eq:chiral-plus-marginal}) in the continuum limit    contain a bound state in spectrum   when the binding potential becomes of the order of the fermion bandwidth.

If we assume that  the   transition line with $\lambda_2<0$ eventually gives rise to a boundary bound state, we can  assemble the  various fixed points into a cohesive phase diagram.  
Restoring the spin index of the fermionic fields, we note that there are actually three bound states, two of which can be occupied if we are to respect the global fermion parity constraint for physical states. Recall that we are dealing with emergent Majorana fermions that stem from the spin fractionalization in the bulk.  The partial occupation of the bound states entails a threefold degeneracy, from which we can construct  a boundary spin-1 operator $\mathbf{S}_{0}=h^{\dagger}_{a}\mathbf{T}^{\phantom\dagger}_{ab}h^{\phantom\dagger}_{b}$. Here $h^\dagger_a$ creates a hole in  the boundary bound state labeled by  $a\in\{1,2,3\}$, and $\mathbf{T}$ is the vector of spin-1 matrices.  Once this effective ``impurity'' spin is formed, the field  theory for the perturbed chiral fixed point allows for another relevant boundary operator with dimension $3/8$:\begin{equation}
H_{\mathcal B}^{\prime}=\lambda_3\mathbf S_0\cdot \sum_\alpha \tr\bm \tau \Phi^{(1/2)}_\alpha(0).
\end{equation}
When $\lambda_3$ flows to strong coupling, it drives the system toward a different  low-energy fixed point than the O fixed points governed by $\lambda_1$.  Importantly, the $\mathbb Z_3$ charge of the complex fermions   forming the bound state is consistent with the chirality of the triplet ground state of the three-spin Hamiltonian in the strong-coupling  limit $J_\chi\gg J'\gg J$, see Eq. (\ref{triplet}). Identifying this strong-coupling limit with the    regime dominated   by $\lambda_3$, we are led to the boundary phase diagram in Fig. \ref{fig:boundary-phase-diagram}.

Coming from the limit of large $J_\chi$, the effective spin 1 at the boundary is coupled to the chains by a Kondo interaction\begin{equation}
\tilde{H}=\tilde{H}_{0}+J_{\mathrm{K}}\mathbf{S}_{0}\cdot\sum_{\alpha}\mathbf{S}_{2,\alpha},
\end{equation}
where $J_{\mathrm{K}}\sim J$ is the   Kondo coupling and $\tilde{H}_{0}$ is the new bulk Hamiltonian where the site $j=1$ of each chain has been removed to form the central spin $\mathbf{S}_{0}$.  The free-moment fixed point  where the spin 1 remains  decoupled   is unstable for $J_{\rm K}>0$ due to the Kondo effect.  At low energies, the boundary   degeneracy must be lifted as the central $S=1$ spin gets overscreened by the symmetric combination  $\mathbf J=\sum_{\alpha}\mathbf J_\alpha$, which defines an SU(2)$_6$ current. To our knowledge, this  low-energy fixed point   has not been discussed in the literature. Formally, the conformally invariant boundary conditions for the stable fixed point can be generated by fusion with the primary fields of the SU(2)$_6$ WZW model once we  identify a suitable conformal embedding. Note that the junction of spin chains has a different  total central charge ($c=9/2$) than its counterpart for  itinerant electrons; thus, the results for the electronic multichannel Kondo model \cite{Affleck1991,Affleck1993} cannot be immediately translated to the spin chain version of the problem. Since our focus here  is on the chiral fixed point, we refrain from  studying this multichannel-Kondo-like fixed point in detail and leave this problem for future work.

%%%%%%%%%%%%%%%%%%%%%%%%%%%%%%%%%%%%%%%%%%%%%%%%%%%%%%%%%%  

\subsection{Comparison with the spin-1/2 case}

At this point it is interesting to contrast our results with those for the junction of spin-1/2 chains discussed in Refs. \cite{Buccheri2018,Buccheri2019}. In the spin-1/2 junction, the low-energy physics of each spin chain is governed by an SU(2)$_1$ WZW model with central charge $c=1$ which is equivalent to a free boson theory \cite{Gogolin1998}. A key difference is that the chiral fixed point is destabilized by only one relevant operator in this case, and   there is no marginal  deformation that would lead to a smoothly varying conductance asymmetry. The unstable chiral fixed point in the spin-1/2 case separates a weak-coupling regime, given by disconnected spin chains, from a strong-coupling regime, dominated by a spin-chain version of the three-channel Kondo fixed point. The latter two phases  find their analogs  in the spin-1 problem, but there is no boundary singlet phase in the spin-1/2 junction.

%%%%%%%%%%%%%%%%%%%%%%%%%%%%%%%%%%%%%%%%%%%%%%%%%%%%%%%%%%

\section{Conclusions}\label{sec:conclusions}

We studied the properties of a chiral junction of  isotropic spin-1 chains with bulk interactions tuned to the critical point between the Haldane phase and the dimerized phase. We argued that a chiral fixed point can be reached by varying the  coupling constants for two boundary interactions in the lattice model. Based on a low-energy effective  field theory, we found that the chiral fixed point is unstable, being perturbed by   a relevant and a marginal boundary operator. We established that the relevant coupling governs a   phase transition between two disconnected regimes, which correspond to trivially decoupled chains and to the formation of a spin-singlet state out of the boundary spins. On the other hand,   the marginal perturbation  produces a phase shift for the emergent fermionic modes and  controls the spin conductance   along the transition line between the decoupled and singlet phases. The formation of a spin-triplet boundary state for a strong three-spin interaction can   be understood through  the formation of  fermion bound states when the marginal coupling  creates a strong binding potential.  Therefore, despite its instability, the perturbed chiral fixed point   governs the low-energy physics of the junction over a   wide parameter regime.

Our analytical predictions can be tested numerically using the methods of Refs. \cite{Guo2006,Rahmani2010,Buccheri2018,Buccheri2019}. One could pinpoint the location of the chiral fixed point by combining the information about the power law  decay of the three-spin correlation, the sign inversion  of the dimerization, and the asymmetry of the spin conductance tensor. The properties   of the strong-coupling regime with  dominant three-spin interaction,  governed by a multichannel-Kondo-like fixed point, remain an open question. As a matter of fact, other nontrivial fixed points may be found in a more general boundary phase diagram that includes a boundary biquadratic interaction.

Our junction is a member of a broad family of quantum spin circulators. If each member of this family is labeled by the spin $S$ of the critical chains, this problem is equivalent to searching for the chiral fixed point in junctions of $\mathrm{SU(2)}_{k=2S}$ WZW models.  In this context, our work serves as the simplest generalization of the chiral fixed point of spin-1/2 chains \cite{Buccheri2018,Buccheri2019} to a higher-level WZW model. The importance of such a generalization can be viewed more directly in the construction of strongly correlated  topological phases in two dimensions. While the $\mathrm{SU(2)}_{1}$ model is related to the construction of a Kalmeyer-Laughlin chiral spin liquid \cite{Ferraz2019,Kalmeyer1987}, an analogous construction for the $\mathrm{SU(2)}_{2}$ model would lead to a non-Abelian chiral spin liquid  with SU(2)$_2$ anyons  \cite{Greiter2009}. This construction will be discussed elsewhere.  More exotic  possibilities, such as parafermionic chiral spin liquids,  are offered by starting from even higher-level $\mathrm{SU(2)}_{k}$ WZW models. However,  the number of relevant boundary operators that can be constructed from the  primary fields $\Phi^{(j)}$ with $j=\frac12, 1, \dots, \frac k2$ in the SU(2)$_k$ WZW  model increases with $k$. The generalization of our  results indicates that it will be necessary to fine tune an increasing number of parameters in the lattice model to reach the chiral fixed point, which should appear as a multicritical point in the boundary phase diagram.

%%%%%%%%%%%%%%%%%%%%%%%%%%%%%%%%%%%%%%%%%%%%%%%%%%%%%%%%%%

\begin{acknowledgments}

We thank Claudio Chamon for many helpful discussions. We are also indebted to Flavia   Ramos for sharing   preliminary numerical results on the junction of spin-1 chains. This paper was supported by the Brazilian funding agencies CAPES (H.B.X.) and CNPq (R.G.P.).  Research at IIP-UFRN is funded by   Brazilian ministries MEC and MCTI. 

\end{acknowledgments}

%%%%%%%%%%%%%%%%%%%%%%%%%%%%%%%%%%%%%%%%%%%%%%%%%%%%%%%%%%

\appendix

%%%%%%%%%%%%%%%%%%%%%%%%%%%%%%%%%%%%%%%%%%%%%%%%%%%%%%%%%%

\section{Bosonization of the Ising model}\label{app:bosonization}

In this appendix, we review the continuum limit formulation of the critical Ising model and establish the bosonization formulas for a theory of two Ising models used in the main text \cite{Shelton1996,Allen1997,Gogolin1998,Allen2000}.

The Hamiltonian of the critical Ising model is that of a free, massless Majorana fermion
\begin{equation}
H=\frac{iv}{2}\int dx \big(\xi\nabla\xi-\bar{\xi}\nabla\bar{\xi}\big).
\end{equation}
This is a conformally invariant theory with central charge $c=1/2$. The nontrivial scaling fields of the theory are the energy operator $\varepsilon$, the order operator $\sigma$, and the disorder operator $\mu$. Their conformal weights $(h,\bar{h})$ are
\begin{equation}
\varepsilon:(\tfrac{1}{2},\tfrac{1}{2}),\qquad
\sigma:(\tfrac{1}{16},\tfrac{1}{16}),\qquad
\mu:(\tfrac{1}{16},\tfrac{1}{16}).
\end{equation}
The order and disorder fields obey the following set of OPEs
\begin{align}
\sigma(z,\bar{z})\sigma(0)&=\frac{1}{\sqrt{2}|z|^{1/4}}+\frac{\pi}{\sqrt{2}}|z|^{3/4}\varepsilon(0)+\cdots,\nonumber\\
\mu(z,\bar{z})\mu(0)&=\frac{1}{\sqrt{2}|z|^{1/4}}-\frac{\pi}{\sqrt{2}}|z|^{3/4}\varepsilon(0)+\cdots,
\end{align}
and
\begin{align}
\mu(z,\bar{z})\sigma(0)&=\sqrt{\frac{\pi}{2}}\frac{e^{i\frac{\pi}{4}}z^{\frac{1}{2}}\xi(0)-e^{-i\frac{\pi}{4}}\bar{z}^{\frac{1}{2}}\bar{\xi}(0)}{|z|^{\frac{1}{4}}}+\cdots,\nonumber\\
\sigma(z,\bar{z})\mu(0)&=\sqrt{\frac{\pi}{2}}\frac{e^{-i\frac{\pi}{4}}z^{\frac{1}{2}}\xi(0)-e^{i\frac{\pi}{4}}\bar{z}^{\frac{1}{2}}\bar{\xi}(0)}{|z|^{\frac{1}{4}}}+\cdots.
\end{align}
This second set of OPEs tell us that the product of order and disorder operators produces a fermion. This means that we must associate a fermionic character to either $\mu$ or $\sigma$ for consistency \cite{Allen2000}. In this paper, we arbitrarily choose $\mu$ to incorporate the anticommutative properties of the fermions. Since $\varepsilon=i\xi\bar{\xi}$, it suffices to examine the OPEs associated to the fusion rules $\xi\times\mu\to\sigma$ and $\xi\times\sigma\to\mu$. For the holomorphic component, we have \begin{align}\label{eq:OPE-xi-times-spin}
\xi(z)\mu(0)&=\frac{e^{i\pi/4}}{\sqrt{4\pi z}}\sigma(0)+\cdots,\nonumber\\
\xi(z)\sigma(0)&=\frac{e^{-i\pi/4}}{\sqrt{4\pi z}}\mu(0)+\cdots.
\end{align}
The OPEs for the right-moving fermions $\bar{\xi}$ are then
\begin{align}\label{eq:OPE-bar-xi-times-spin}
\bar{\xi}(\bar{z})\mu(0)&=-\frac{e^{-i\pi/4}}{\sqrt{4\pi \bar{z}}}\sigma(0)+\cdots,\nonumber\\
\bar{\xi}(\bar{z})\sigma(0)&=-\frac{e^{i\pi/4}}{\sqrt{4\pi \bar{z}}}\mu(0)+\cdots.
\end{align}
These coefficients differ from the ones found in the textbooks by Mussardo \cite{Mussardo2010} and Di Francesco {\it et al.} \cite{DiFrancesco1997}. However, as pointed out by Allen and S\'en\'echal in Ref. \cite{Allen1997}, there is some arbitrariness in these constants because of the nonlocal character of these operators. We stress that our notation is  consistent with the bosonization formulas presented below.

Let us now establish the bosonization formulas \cite{Gogolin1998}. To begin with, we introduce the chiral boson fields $\phi$ and $\bar{\phi}$, which are uniquely defined by their two-point functions
\begin{equation}
\vev{\phi(z)\phi(0)}=-\frac{1}{4}\ln z,\quad 
\vev{\bar{\phi}(\bar{z})\bar{\phi}(0)}=-\frac{1}{4}\ln \bar{z},
\end{equation}
with $\vev{\phi(z)\bar{\phi}(\bar{z})}=0$. In terms of these, the bosonization formulas are
\begin{equation}\label{eq:bosonization-formula-app}
\xi^{1}+i\xi^{2}=\frac{1}{\sqrt{\pi}}e^{i2\phi},\qquad
\bar{\xi}^{\,1}+i\bar{\xi}^{\,2}=\frac{1}{\sqrt{\pi}}e^{i2\bar{\phi}}.
\end{equation}
To ensure the anticommutation of the fermions, we impose the equal-time commutators
\begin{align}
\comm{\phi(x)}{\phi(x')}&=-i\frac{\pi}{4}\sgn(x-x'),\nonumber\\
\comm{\bar{\phi}(x)}{\bar{\phi}(x')}&=i\frac{\pi}{4}\sgn(x-x'),\nonumber\\
\comm{\phi(x)}{\bar{\phi}(x')}&=-i\frac{\pi}{4}.
\end{align}
Here $\sgn (x)$ stands for the sign function with $\sgn(0)=0$. From Eq. (\ref{eq:bosonization-formula-app}) it follows that the currents are given by
\begin{equation}
J^{z}=-i\xi^{1}\xi^{2}=\frac{i}{\pi}\partial\phi,\qquad
\bar{J}^{z}=-i\bar{\xi}^{\,1}\bar{\xi}^{\,2}=\frac{i}{\pi}\bar{\partial}\bar{\phi},
\end{equation}
where $\partial$ and $\bar \partial$ denote the derivatives with respect to $z$ and $\bar z$, respectively. 
Here we used the definition in Eq. (\ref{eq:spin-currents}). If we now introduce the scalar field $\varphi=\phi-\bar{\phi}$ and its dual $\theta=\phi+\bar{\phi}$, we can write
\begin{equation}
J^{z}+\bar{J}^{z}=\frac{1}{\pi}\nabla\varphi,\qquad
J^{z}-\bar{J}^{z}=\frac{1}{\pi}\nabla\theta.
\end{equation}
In the same token, to find the bosonization formulas for the energy operators, we   multiply the normal-ordered exponentials in Eq. (\ref{eq:bosonization-formula-app}) and obtain 
\begin{equation}
\varepsilon^{1}+\varepsilon^{2}=\frac{1}{\pi}\cos2\varphi,\qquad
\varepsilon^{1}-\varepsilon^{2}=-\frac{1}{\pi}\cos2\theta.
\end{equation}
To express order and disorder operators, we consider products of operators such as $\sigma^{1}\sigma^{2}$. The products of two order and two disorder operators are given by 
\begin{equation}\label{eq:bosonization-doubled-order-disorder}
\sigma^{1}\sigma^{2}=\cos\varphi,\qquad i\mu^{1}\mu^{2}=\sin\varphi,
\end{equation}
and the mixed products are
\begin{equation}\label{eq:bosonization-mixed-order-disorder}
\mu^{1}\sigma^{2}=\cos\theta,\qquad \sigma^{1}\mu^{2}=\sin\theta.
\end{equation}
One way to find the correct correspondence of these formulas is to ensure that the perturbations $m\int dx(\varepsilon^{1}\pm\varepsilon^{2})$ to the effective Hamiltonian induce the correct phase transitions for positive and negative $m$, cf. Ref. \cite{Shelton1996}. For example,  $m>0$ in $m\int dx(\varepsilon^{1}+\varepsilon^{2})$ corresponds to the disordered phase of the Ising models. In the boson formulation, this can be interpreted as pinning $\varphi\to\pm\pi/2$ to minimize the energy, which according to Eq. (\ref{eq:bosonization-doubled-order-disorder}) yields $\vev{\sigma^{a}}=0$ and $\vev{\mu^{a}}\neq0$, as it should be.

%%%%%%%%%%%%%%%%%%%%%%%%%%%%%%%%%%%%%%%%%%%%%%%%%%%%%%%%%%

\section{Bound state of chiral fermions}\label{app:bound-state}

In this appendix, we examine  lattice realizations   for the problem of a chiral fermion scattering off   a delta-function potential   in Eq. (\ref{eq:chiral-plus-marginal}). Our goal   is to show that on a regularized platform a bound state can be formed when the binding potential is strong enough.

\begin{figure}
\centering
\includegraphics[width=.9\linewidth]{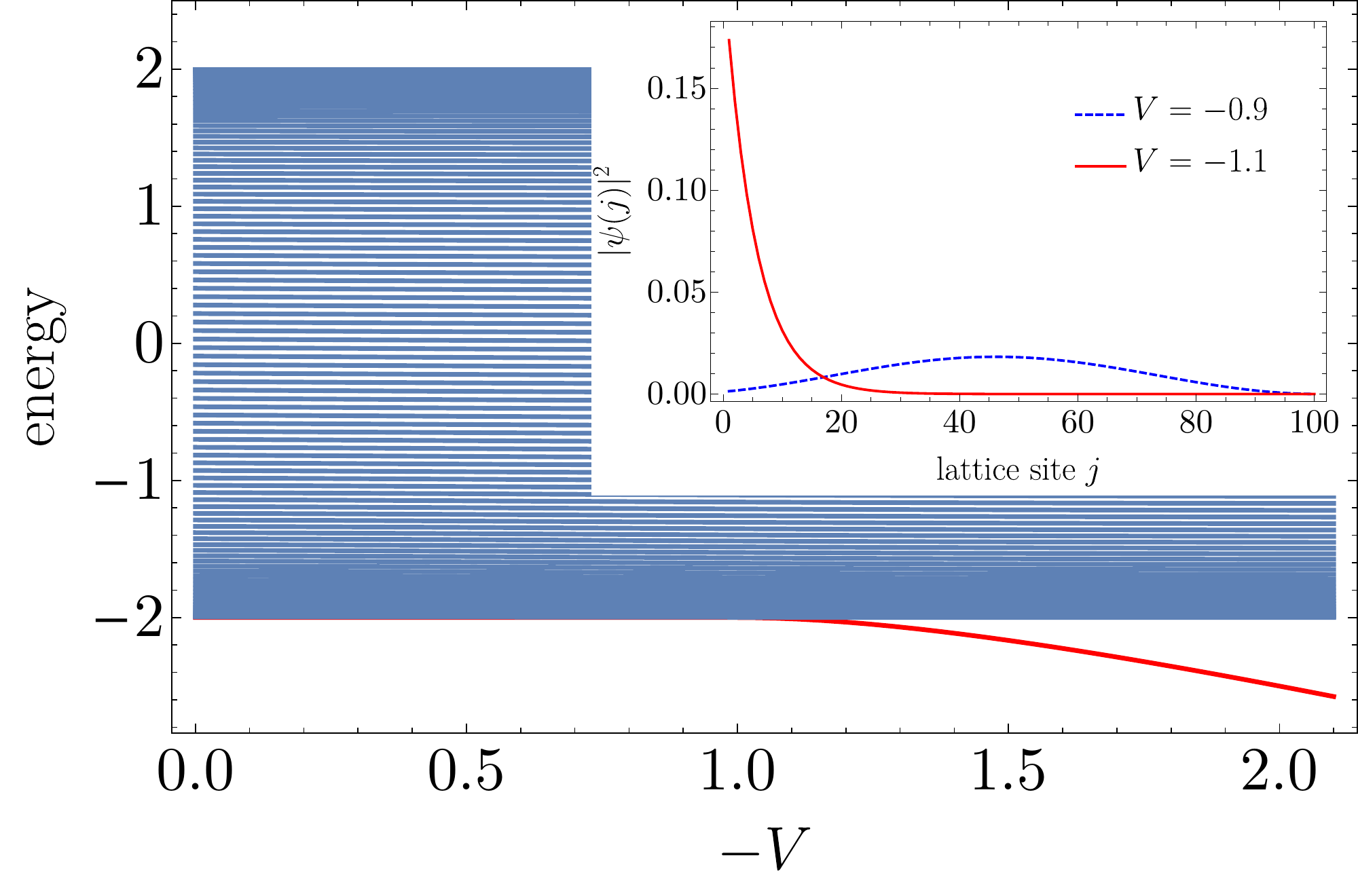}
\caption{Energy spectrum as a function of the impurity potential for a chain with $N=100$ sites, see Eq. (\ref{eq:microscopic-model}). The lowest-energy state detaches  from the continuum   for $V\lesssim-1$, indicating a bound state has been formed. Inset: Probability density of the lowest-energy state for $V=-0.9$ and $V=-1.1$.}
\label{fig:spectrum-lattice-inset}
\end{figure}

First,   consider a tight-binding model  of spinless fermions hopping on an open  chain with $N$ sites. We then  add an onsite potential $V$ acting at the first site. The Hamiltonian reads  
\begin{equation}\label{eq:microscopic-model}
H=-\sum_{j=1}^{N-1}\big(c_{j}^{\dagger}c_{j+1}^{\phantom\dagger}+\hc\big)+Vc_{1}^{\dagger}c_{1}^{\phantom\dagger},
\end{equation}
where $\acomm{c_{j}^{\phantom\dagger}}{c_{j'}^{\dagger}}=\delta_{jj'}$ and the hopping parameter has been set to unity. Assuming that $V$ is weak compared with the Fermi energy, we can take the continuum limit starting from the model with $V=0$ and then include  the local potential as a perturbation. In the limit $N\to\infty$, we   expand the fermion operator around the Fermi points $\pm k_{F}$ as 
\begin{equation}\label{eq:fermion-cont-representation}
c_{j}\sim\psi(x)e^{-ik_{F}x}+\bar{\psi}(x)e^{ik_{F}x}.
\end{equation}
When expressed in terms of the slow fields $\psi$ and $\bar{\psi}$, the free Hamiltonian takes the form
\begin{equation}
H_{0}\simeq iv_{F}\int_{0}^{\infty}dx\big(\psi^{\dagger}\nabla\psi-\bar{\psi}^{\dagger}\nabla\bar{\psi}\big),
\end{equation}
with $v_{F}=2\sin k_{F}$ the Fermi velocity. The open boundary at $x=0$ implies that  $\bar{\psi}$ and $\psi$ are not independent. Imposing that   $c_0=0$, we obtain  $\bar{\psi}(0)=-\psi(0)$. We can then write $\bar\psi$ as the analytic continuation of $\psi$ to the negative-$x$ axis:\begin{equation}\label{eq:fermion-open-BC}
\bar{\psi}(x)=-\psi(-x).
\end{equation}
Using this relation, we unfold the space interval and write $H_{0}$ only in terms of $\psi$ in the form\begin{equation}
H_{0}= iv_{F}\int_{-\infty}^{\infty}dx\,\psi^{\dagger}\nabla\psi.
\end{equation}
We now include  the local potential $H_{I}=Vc_{1}^{\dagger}c_{1}^{\phantom\dagger}$. Its continuum representation follows directly from Eqs. (\ref{eq:fermion-cont-representation}) and (\ref{eq:fermion-open-BC}). The result   is
\begin{equation}
H_{I}\simeq v_{F}\gamma\,\psi^{\dagger}(0)\psi(0),
\end{equation}
where $\gamma=2V\sin k_{F}$. Setting $v_F=1$, we can write  the effective Hamiltonian  $H=H_{0}+H_{I}$   as
\begin{equation}
H=\int dx\,\psi^{\dagger}\big[i\nabla+\gamma\delta(x)\big]\psi.\label{lineardisp}
\end{equation}
Thus, the chain of spinless fermions in Eq. (\ref{eq:microscopic-model}) indeed works as a microscopic model for the problem in Eq. (\ref{eq:chiral-plus-marginal}).

As discussed in Sec. \ref{sec:bound}, the low-energy effective  Hamiltonian with linearized dispersion in Eq. (\ref{lineardisp}) does not predict the formation of a bound state for $\gamma<0$.  However, we can now directly analyze the single-particle spectrum for  the lattice model  of Eq. (\ref{eq:microscopic-model}). We have diagonalized this  Hamiltonian numerically for a chain with $N=100$ sites. As shown in Fig. \ref{fig:spectrum-lattice-inset}, we find a bound state below the continuum of extended states for $V\lesssim-1$. The inset of  Fig. \ref{fig:spectrum-lattice-inset} confirms  that the wave function of this state is  localized near the chain end. This result shows that the formation of a bound state is indeed beyond the scope of our low-energy theory. 

The finite critical value of $|V|$ required for the formation of a bound state in the model of Eq. (\ref{eq:microscopic-model})  can be interpreted as follows. The bound state appears below the lower threshold of the continuum of extended states. This lower threshold is defined by states with momentum $k\to0$, whose wave function vanishes near the boundary at $x=0$. As a result, the states at the bottom of the band are rather insensitive to a weak  scattering potential at the boundary. The attractive potential only gives rise to a bound state when  it becomes comparable with the bandwidth, $|V|\sim 1$. 

One may object that the model of Eq. (\ref{eq:microscopic-model})  only describes a chiral fermion at low energies after we use the folding trick of Eq. (\ref{eq:fermion-open-BC}). A single  chiral fermion is not possible in a 1D lattice model because of the fermion doubling problem. For this reason, we have also considered a model where the chiral fermion is realized  as the edge state  of a two-dimensional topological phase. Here we consider the Haldane model for a Chern insulator on the honeycomb lattice \cite{Haldane1988}. We diagonalize the Hamiltonian numerically on a finite cylinder with zigzag  edges, including the on-site potential $V$ at one boundary site. In this case the critical value of $|V|$ above which the bound state arises depends on the magnitude of the imaginary second-neighbor hopping parameter. Nevertheless, as  a general and robust result, we observe that there is no  bound state at weak coupling. The bound state only appears below the continuum of  two-dimensional extended states  when $|V|$ becomes of the order of the bandwidth. 

In summary, the formation of the bound state of   chiral fermions in the presence of a delta-function potential is a nonuniversal phenomenon which depends on information about the high-energy spectrum. In this sense, it is not surprising that it is not captured by our low-energy effective field theory.

%%%%%%%%%%%%%%%%%%%%%%%%%%%%%%%%%%%%%%%%%%%%%%%%%%%%%%%%%%

\bibliography{references}

\begin{thebibliography}{57}%
\makeatletter
\providecommand \@ifxundefined [1]{%
 \@ifx{#1\undefined}
}%
\providecommand \@ifnum [1]{%
 \ifnum #1\expandafter \@firstoftwo
 \else \expandafter \@secondoftwo
 \fi
}%
\providecommand \@ifx [1]{%
 \ifx #1\expandafter \@firstoftwo
 \else \expandafter \@secondoftwo
 \fi
}%
\providecommand \natexlab [1]{#1}%
\providecommand \enquote  [1]{``#1''}%
\providecommand \bibnamefont  [1]{#1}%
\providecommand \bibfnamefont [1]{#1}%
\providecommand \citenamefont [1]{#1}%
\providecommand \href@noop [0]{\@secondoftwo}%
\providecommand \href [0]{\begingroup \@sanitize@url \@href}%
\providecommand \@href[1]{\@@startlink{#1}\@@href}%
\providecommand \@@href[1]{\endgroup#1\@@endlink}%
\providecommand \@sanitize@url [0]{\catcode `\\12\catcode `\$12\catcode
  `\&12\catcode `\#12\catcode `\^12\catcode `\_12\catcode `\%12\relax}%
\providecommand \@@startlink[1]{}%
\providecommand \@@endlink[0]{}%
\providecommand \url  [0]{\begingroup\@sanitize@url \@url }%
\providecommand \@url [1]{\endgroup\@href {#1}{\urlprefix }}%
\providecommand \urlprefix  [0]{URL }%
\providecommand \Eprint [0]{\href }%
\providecommand \doibase [0]{https://doi.org/}%
\providecommand \selectlanguage [0]{\@gobble}%
\providecommand \bibinfo  [0]{\@secondoftwo}%
\providecommand \bibfield  [0]{\@secondoftwo}%
\providecommand \translation [1]{[#1]}%
\providecommand \BibitemOpen [0]{}%
\providecommand \bibitemStop [0]{}%
\providecommand \bibitemNoStop [0]{.\EOS\space}%
\providecommand \EOS [0]{\spacefactor3000\relax}%
\providecommand \BibitemShut  [1]{\csname bibitem#1\endcsname}%
\let\auto@bib@innerbib\@empty
%</preamble>
\bibitem [{\citenamefont {Alicea}\ \emph {et~al.}(2011)\citenamefont {Alicea},
  \citenamefont {Oreg}, \citenamefont {Refael}, \citenamefont {von Oppen},\
  and\ \citenamefont {Fisher}}]{Alicea2011}%
  \BibitemOpen
  \bibfield  {author} {\bibinfo {author} {\bibfnamefont {J.}~\bibnamefont
  {Alicea}}, \bibinfo {author} {\bibfnamefont {Y.}~\bibnamefont {Oreg}},
  \bibinfo {author} {\bibfnamefont {G.}~\bibnamefont {Refael}}, \bibinfo
  {author} {\bibfnamefont {F.}~\bibnamefont {von Oppen}},\ and\ \bibinfo
  {author} {\bibfnamefont {M.~P.~A.}\ \bibnamefont {Fisher}},\ }\bibfield
  {title} {\bibinfo {title} {{Non-Abelian statistics and topological quantum
  information processing in 1D wire networks}},\ }\href
  {https://doi.org/10.1038/nphys1915} {\bibfield  {journal} {\bibinfo
  {journal} {Nat. Phys.}\ }\textbf {\bibinfo {volume} {7}},\ \bibinfo {pages}
  {412} (\bibinfo {year} {2011})}\BibitemShut {NoStop}%
\bibitem [{\citenamefont {Qiu}\ \emph {et~al.}(2020)\citenamefont {Qiu},
  \citenamefont {Zoller},\ and\ \citenamefont {Li}}]{Qiu2020}%
  \BibitemOpen
  \bibfield  {author} {\bibinfo {author} {\bibfnamefont {X.}~\bibnamefont
  {Qiu}}, \bibinfo {author} {\bibfnamefont {P.}~\bibnamefont {Zoller}},\ and\
  \bibinfo {author} {\bibfnamefont {X.}~\bibnamefont {Li}},\ }\bibfield
  {title} {\bibinfo {title} {{Programmable Quantum Annealing Architectures with
  Ising Quantum Wires}},\ }\href {https://doi.org/10.1103/PRXQuantum.1.020311}
  {\bibfield  {journal} {\bibinfo  {journal} {PRX Quantum}\ }\textbf {\bibinfo
  {volume} {1}},\ \bibinfo {pages} {020311} (\bibinfo {year}
  {2020})}\BibitemShut {NoStop}%
\bibitem [{\citenamefont {Medina}\ \emph {et~al.}(2013)\citenamefont {Medina},
  \citenamefont {Green},\ and\ \citenamefont {Chamon}}]{Medina2013}%
  \BibitemOpen
  \bibfield  {author} {\bibinfo {author} {\bibfnamefont {J.}~\bibnamefont
  {Medina}}, \bibinfo {author} {\bibfnamefont {D.}~\bibnamefont {Green}},\ and\
  \bibinfo {author} {\bibfnamefont {C.}~\bibnamefont {Chamon}},\ }\bibfield
  {title} {\bibinfo {title} {{Networks of quantum wire junctions: A system with
  quantized integer Hall resistance without vanishing longitudinal
  resistivity}},\ }\href {https://doi.org/10.1103/PhysRevB.87.045128}
  {\bibfield  {journal} {\bibinfo  {journal} {Phys. Rev. B}\ }\textbf {\bibinfo
  {volume} {87}},\ \bibinfo {pages} {045128} (\bibinfo {year}
  {2013})}\BibitemShut {NoStop}%
\bibitem [{\citenamefont {San-Jose}\ and\ \citenamefont
  {Prada}(2013)}]{SanJose2013}%
  \BibitemOpen
  \bibfield  {author} {\bibinfo {author} {\bibfnamefont {P.}~\bibnamefont
  {San-Jose}}\ and\ \bibinfo {author} {\bibfnamefont {E.}~\bibnamefont
  {Prada}},\ }\bibfield  {title} {\bibinfo {title} {Helical networks in twisted
  bilayer graphene under interlayer bias},\ }\href
  {https://doi.org/10.1103/PhysRevB.88.121408} {\bibfield  {journal} {\bibinfo
  {journal} {Phys. Rev. B}\ }\textbf {\bibinfo {volume} {88}},\ \bibinfo
  {pages} {121408} (\bibinfo {year} {2013})}\BibitemShut {NoStop}%
\bibitem [{\citenamefont {Ferraz}\ \emph {et~al.}(2019)\citenamefont {Ferraz},
  \citenamefont {Ramos}, \citenamefont {Egger},\ and\ \citenamefont
  {Pereira}}]{Ferraz2019}%
  \BibitemOpen
  \bibfield  {author} {\bibinfo {author} {\bibfnamefont {G.}~\bibnamefont
  {Ferraz}}, \bibinfo {author} {\bibfnamefont {F.~B.}\ \bibnamefont {Ramos}},
  \bibinfo {author} {\bibfnamefont {R.}~\bibnamefont {Egger}},\ and\ \bibinfo
  {author} {\bibfnamefont {R.~G.}\ \bibnamefont {Pereira}},\ }\bibfield
  {title} {\bibinfo {title} {{Spin chain network construction of chiral spin
  liquids}},\ }\href {https://doi.org/10.1103/PhysRevLett.123.137202}
  {\bibfield  {journal} {\bibinfo  {journal} {Phys. Rev. Lett.}\ }\textbf
  {\bibinfo {volume} {123}},\ \bibinfo {pages} {137202} (\bibinfo {year}
  {2019})}\BibitemShut {NoStop}%
\bibitem [{\citenamefont {Lee}\ \emph {et~al.}(2021)\citenamefont {Lee},
  \citenamefont {Oshikawa},\ and\ \citenamefont {Cho}}]{Lee2021}%
  \BibitemOpen
  \bibfield  {author} {\bibinfo {author} {\bibfnamefont {J.~M.}\ \bibnamefont
  {Lee}}, \bibinfo {author} {\bibfnamefont {M.}~\bibnamefont {Oshikawa}},\ and\
  \bibinfo {author} {\bibfnamefont {G.~Y.}\ \bibnamefont {Cho}},\ }\bibfield
  {title} {\bibinfo {title} {{Non-Fermi Liquids in Conducting Two-Dimensional
  Networks}},\ }\href {https://doi.org/10.1103/PhysRevLett.126.186601}
  {\bibfield  {journal} {\bibinfo  {journal} {Phys. Rev. Lett.}\ }\textbf
  {\bibinfo {volume} {126}},\ \bibinfo {pages} {186601} (\bibinfo {year}
  {2021})}\BibitemShut {NoStop}%
\bibitem [{\citenamefont {Chou}\ \emph {et~al.}(2021)\citenamefont {Chou},
  \citenamefont {Wu},\ and\ \citenamefont {Sau}}]{Chou2021}%
  \BibitemOpen
  \bibfield  {author} {\bibinfo {author} {\bibfnamefont {Y.-Z.}\ \bibnamefont
  {Chou}}, \bibinfo {author} {\bibfnamefont {F.}~\bibnamefont {Wu}},\ and\
  \bibinfo {author} {\bibfnamefont {J.~D.}\ \bibnamefont {Sau}},\ }\bibfield
  {title} {\bibinfo {title} {Charge density wave and finite-temperature
  transport in minimally twisted bilayer graphene},\ }\href
  {https://doi.org/10.1103/PhysRevB.104.045146} {\bibfield  {journal} {\bibinfo
   {journal} {Phys. Rev. B}\ }\textbf {\bibinfo {volume} {104}},\ \bibinfo
  {pages} {045146} (\bibinfo {year} {2021})}\BibitemShut {NoStop}%
\bibitem [{\citenamefont {Cardy}(1984)}]{Cardy1984}%
  \BibitemOpen
  \bibfield  {author} {\bibinfo {author} {\bibfnamefont {J.~L.}\ \bibnamefont
  {Cardy}},\ }\bibfield  {title} {\bibinfo {title} {Conformal invariance and
  surface critical behavior},\ }\href
  {https://doi.org/https://doi.org/10.1016/0550-3213(84)90241-4} {\bibfield
  {journal} {\bibinfo  {journal} {Nucl. Phys. B}\ }\textbf {\bibinfo {volume}
  {240}},\ \bibinfo {pages} {514} (\bibinfo {year} {1984})}\BibitemShut
  {NoStop}%
\bibitem [{\citenamefont {Cardy}\ and\ \citenamefont
  {Lewellen}(1991)}]{Cardy1991}%
  \BibitemOpen
  \bibfield  {author} {\bibinfo {author} {\bibfnamefont {J.~L.}\ \bibnamefont
  {Cardy}}\ and\ \bibinfo {author} {\bibfnamefont {D.~C.}\ \bibnamefont
  {Lewellen}},\ }\bibfield  {title} {\bibinfo {title} {Bulk and boundary
  operators in conformal field theory},\ }\href
  {https://doi.org/https://doi.org/10.1016/0370-2693(91)90828-E} {\bibfield
  {journal} {\bibinfo  {journal} {Phys. Lett. B}\ }\textbf {\bibinfo {volume}
  {259}},\ \bibinfo {pages} {274} (\bibinfo {year} {1991})}\BibitemShut
  {NoStop}%
\bibitem [{\citenamefont {Affleck}\ and\ \citenamefont
  {Ludwig}(1991)}]{Affleck1991}%
  \BibitemOpen
  \bibfield  {author} {\bibinfo {author} {\bibfnamefont {I.}~\bibnamefont
  {Affleck}}\ and\ \bibinfo {author} {\bibfnamefont {A.~W.}\ \bibnamefont
  {Ludwig}},\ }\bibfield  {title} {\bibinfo {title} {{Critical theory of
  overscreened Kondo fixed points}},\ }\href
  {https://doi.org/https://doi.org/10.1016/0550-3213(91)90419-X} {\bibfield
  {journal} {\bibinfo  {journal} {Nucl. Phys. B}\ }\textbf {\bibinfo {volume}
  {360}},\ \bibinfo {pages} {641} (\bibinfo {year} {1991})}\BibitemShut
  {NoStop}%
\bibitem [{\citenamefont {Affleck}\ and\ \citenamefont
  {Ludwig}(1993)}]{Affleck1993}%
  \BibitemOpen
  \bibfield  {author} {\bibinfo {author} {\bibfnamefont {I.}~\bibnamefont
  {Affleck}}\ and\ \bibinfo {author} {\bibfnamefont {A.~W.~W.}\ \bibnamefont
  {Ludwig}},\ }\bibfield  {title} {\bibinfo {title} {{Exact
  conformal-field-theory results on the multichannel Kondo effect:
  Single-fermion Green's function, self-energy, and resistivity}},\ }\href
  {https://doi.org/10.1103/PhysRevB.48.7297} {\bibfield  {journal} {\bibinfo
  {journal} {Phys. Rev. B}\ }\textbf {\bibinfo {volume} {48}},\ \bibinfo
  {pages} {7297} (\bibinfo {year} {1993})}\BibitemShut {NoStop}%
\bibitem [{\citenamefont {Vojta}(2006)}]{Vojta2006}%
  \BibitemOpen
  \bibfield  {author} {\bibinfo {author} {\bibfnamefont {M.}~\bibnamefont
  {Vojta}},\ }\bibfield  {title} {\bibinfo {title} {Impurity quantum phase
  transitions},\ }\href {https://doi.org/10.1080/14786430500070396} {\bibfield
  {journal} {\bibinfo  {journal} {Philos. Mag.}\ }\textbf {\bibinfo {volume}
  {86}},\ \bibinfo {pages} {1807} (\bibinfo {year} {2006})}\BibitemShut
  {NoStop}%
\bibitem [{\citenamefont {Kane}\ and\ \citenamefont
  {Fisher}(1992{\natexlab{a}})}]{Kane1992a}%
  \BibitemOpen
  \bibfield  {author} {\bibinfo {author} {\bibfnamefont {C.~L.}\ \bibnamefont
  {Kane}}\ and\ \bibinfo {author} {\bibfnamefont {M.~P.~A.}\ \bibnamefont
  {Fisher}},\ }\bibfield  {title} {\bibinfo {title} {{Transport in a
  one-channel Luttinger liquid}},\ }\href
  {https://doi.org/10.1103/PhysRevLett.68.1220} {\bibfield  {journal} {\bibinfo
   {journal} {Phys. Rev. Lett.}\ }\textbf {\bibinfo {volume} {68}},\ \bibinfo
  {pages} {1220} (\bibinfo {year} {1992}{\natexlab{a}})}\BibitemShut {NoStop}%
\bibitem [{\citenamefont {Kane}\ and\ \citenamefont
  {Fisher}(1992{\natexlab{b}})}]{Kane1992b}%
  \BibitemOpen
  \bibfield  {author} {\bibinfo {author} {\bibfnamefont {C.~L.}\ \bibnamefont
  {Kane}}\ and\ \bibinfo {author} {\bibfnamefont {M.~P.~A.}\ \bibnamefont
  {Fisher}},\ }\bibfield  {title} {\bibinfo {title} {Transmission through
  barriers and resonant tunneling in an interacting one-dimensional electron
  gas},\ }\href {https://doi.org/10.1103/PhysRevB.46.15233} {\bibfield
  {journal} {\bibinfo  {journal} {Phys. Rev. B}\ }\textbf {\bibinfo {volume}
  {46}},\ \bibinfo {pages} {15233} (\bibinfo {year}
  {1992}{\natexlab{b}})}\BibitemShut {NoStop}%
\bibitem [{\citenamefont {Chamon}\ \emph {et~al.}(2003)\citenamefont {Chamon},
  \citenamefont {Oshikawa},\ and\ \citenamefont {Affleck}}]{Chamon2003}%
  \BibitemOpen
  \bibfield  {author} {\bibinfo {author} {\bibfnamefont {C.}~\bibnamefont
  {Chamon}}, \bibinfo {author} {\bibfnamefont {M.}~\bibnamefont {Oshikawa}},\
  and\ \bibinfo {author} {\bibfnamefont {I.}~\bibnamefont {Affleck}},\
  }\bibfield  {title} {\bibinfo {title} {{Junctions of three quantum wires and
  the dissipative Hofstadter model}},\ }\href
  {https://doi.org/10.1103/PhysRevLett.91.206403} {\bibfield  {journal}
  {\bibinfo  {journal} {Phys. Rev. Lett.}\ }\textbf {\bibinfo {volume} {91}},\
  \bibinfo {pages} {206403} (\bibinfo {year} {2003})}\BibitemShut {NoStop}%
\bibitem [{\citenamefont {Barnab\'e-Th\'eriault}\ \emph
  {et~al.}(2005)\citenamefont {Barnab\'e-Th\'eriault}, \citenamefont {Sedeki},
  \citenamefont {Meden},\ and\ \citenamefont {Sch\"onhammer}}]{Barnabe2005}%
  \BibitemOpen
  \bibfield  {author} {\bibinfo {author} {\bibfnamefont {X.}~\bibnamefont
  {Barnab\'e-Th\'eriault}}, \bibinfo {author} {\bibfnamefont {A.}~\bibnamefont
  {Sedeki}}, \bibinfo {author} {\bibfnamefont {V.}~\bibnamefont {Meden}},\ and\
  \bibinfo {author} {\bibfnamefont {K.}~\bibnamefont {Sch\"onhammer}},\
  }\bibfield  {title} {\bibinfo {title} {Junction of three quantum wires:
  Restoring time-reversal symmetry by interaction},\ }\href
  {https://doi.org/10.1103/PhysRevLett.94.136405} {\bibfield  {journal}
  {\bibinfo  {journal} {Phys. Rev. Lett.}\ }\textbf {\bibinfo {volume} {94}},\
  \bibinfo {pages} {136405} (\bibinfo {year} {2005})}\BibitemShut {NoStop}%
\bibitem [{\citenamefont {Oshikawa}\ \emph {et~al.}(2006)\citenamefont
  {Oshikawa}, \citenamefont {Chamon},\ and\ \citenamefont
  {Affleck}}]{Oshikawa2006}%
  \BibitemOpen
  \bibfield  {author} {\bibinfo {author} {\bibfnamefont {M.}~\bibnamefont
  {Oshikawa}}, \bibinfo {author} {\bibfnamefont {C.}~\bibnamefont {Chamon}},\
  and\ \bibinfo {author} {\bibfnamefont {I.}~\bibnamefont {Affleck}},\
  }\bibfield  {title} {\bibinfo {title} {Junctions of three quantum wires},\
  }\href {https://doi.org/10.1088/1742-5468/2006/02/p02008} {\bibfield
  {journal} {\bibinfo  {journal} {J. Stat. Mech.: Theor. Exp.}\ }\textbf
  {\bibinfo {volume} {2006}},\ \bibinfo {pages} {P02008} (\bibinfo {year}
  {2006})}\BibitemShut {NoStop}%
\bibitem [{\citenamefont {Eggert}\ and\ \citenamefont
  {Affleck}(1992)}]{Eggert1992}%
  \BibitemOpen
  \bibfield  {author} {\bibinfo {author} {\bibfnamefont {S.}~\bibnamefont
  {Eggert}}\ and\ \bibinfo {author} {\bibfnamefont {I.}~\bibnamefont
  {Affleck}},\ }\bibfield  {title} {\bibinfo {title} {{Magnetic impurities in
  half-integer-spin Heisenberg antiferromagnetic chains}},\ }\href
  {https://doi.org/10.1103/PhysRevB.46.10866} {\bibfield  {journal} {\bibinfo
  {journal} {Phys. Rev. B}\ }\textbf {\bibinfo {volume} {46}},\ \bibinfo
  {pages} {10866} (\bibinfo {year} {1992})}\BibitemShut {NoStop}%
\bibitem [{\citenamefont {Bayat}\ \emph {et~al.}(2012)\citenamefont {Bayat},
  \citenamefont {Bose}, \citenamefont {Sodano},\ and\ \citenamefont
  {Johannesson}}]{Bayat2012}%
  \BibitemOpen
  \bibfield  {author} {\bibinfo {author} {\bibfnamefont {A.}~\bibnamefont
  {Bayat}}, \bibinfo {author} {\bibfnamefont {S.}~\bibnamefont {Bose}},
  \bibinfo {author} {\bibfnamefont {P.}~\bibnamefont {Sodano}},\ and\ \bibinfo
  {author} {\bibfnamefont {H.}~\bibnamefont {Johannesson}},\ }\bibfield
  {title} {\bibinfo {title} {{Entanglement Probe of Two-Impurity Kondo Physics
  in a Spin Chain}},\ }\href {https://doi.org/10.1103/PhysRevLett.109.066403}
  {\bibfield  {journal} {\bibinfo  {journal} {Phys. Rev. Lett.}\ }\textbf
  {\bibinfo {volume} {109}},\ \bibinfo {pages} {066403} (\bibinfo {year}
  {2012})}\BibitemShut {NoStop}%
\bibitem [{\citenamefont {Bayat}\ \emph {et~al.}(2014)\citenamefont {Bayat},
  \citenamefont {Johannesson}, \citenamefont {Bose},\ and\ \citenamefont
  {Sodano}}]{Bayat2014}%
  \BibitemOpen
  \bibfield  {author} {\bibinfo {author} {\bibfnamefont {A.}~\bibnamefont
  {Bayat}}, \bibinfo {author} {\bibfnamefont {H.}~\bibnamefont {Johannesson}},
  \bibinfo {author} {\bibfnamefont {S.}~\bibnamefont {Bose}},\ and\ \bibinfo
  {author} {\bibfnamefont {P.}~\bibnamefont {Sodano}},\ }\bibfield  {title}
  {\bibinfo {title} {An order parameter for impurity systems at quantum
  criticality},\ }\href {https://doi.org/10.1038/ncomms4784} {\bibfield
  {journal} {\bibinfo  {journal} {Nat. Commun.}\ }\textbf {\bibinfo {volume}
  {5}},\ \bibinfo {pages} {3784} (\bibinfo {year} {2014})}\BibitemShut
  {NoStop}%
\bibitem [{\citenamefont {Tsvelik}(2013)}]{Tsvelik2013}%
  \BibitemOpen
  \bibfield  {author} {\bibinfo {author} {\bibfnamefont {A.~M.}\ \bibnamefont
  {Tsvelik}},\ }\bibfield  {title} {\bibinfo {title} {{Majorana fermion
  realization of a two-channel Kondo effect in a junction of three quantum
  Ising chains}},\ }\href {https://doi.org/10.1103/PhysRevLett.110.147202}
  {\bibfield  {journal} {\bibinfo  {journal} {Phys. Rev. Lett.}\ }\textbf
  {\bibinfo {volume} {110}},\ \bibinfo {pages} {147202} (\bibinfo {year}
  {2013})}\BibitemShut {NoStop}%
\bibitem [{\citenamefont {Giuliano}\ \emph {et~al.}(2016)\citenamefont
  {Giuliano}, \citenamefont {Sodano}, \citenamefont {Tagliacozzo},\ and\
  \citenamefont {Trombettoni}}]{Giuliano2016}%
  \BibitemOpen
  \bibfield  {author} {\bibinfo {author} {\bibfnamefont {D.}~\bibnamefont
  {Giuliano}}, \bibinfo {author} {\bibfnamefont {P.}~\bibnamefont {Sodano}},
  \bibinfo {author} {\bibfnamefont {A.}~\bibnamefont {Tagliacozzo}},\ and\
  \bibinfo {author} {\bibfnamefont {A.}~\bibnamefont {Trombettoni}},\
  }\bibfield  {title} {\bibinfo {title} {{From four- to two-channel Kondo
  effect in junctions of XY spin chains}},\ }\href
  {https://doi.org/https://doi.org/10.1016/j.nuclphysb.2016.05.003} {\bibfield
  {journal} {\bibinfo  {journal} {Nucl. Phys. B}\ }\textbf {\bibinfo {volume}
  {909}},\ \bibinfo {pages} {135} (\bibinfo {year} {2016})}\BibitemShut
  {NoStop}%
\bibitem [{\citenamefont {Buccheri}\ \emph {et~al.}(2018)\citenamefont
  {Buccheri}, \citenamefont {Egger}, \citenamefont {Pereira},\ and\
  \citenamefont {Ramos}}]{Buccheri2018}%
  \BibitemOpen
  \bibfield  {author} {\bibinfo {author} {\bibfnamefont {F.}~\bibnamefont
  {Buccheri}}, \bibinfo {author} {\bibfnamefont {R.}~\bibnamefont {Egger}},
  \bibinfo {author} {\bibfnamefont {R.~G.}\ \bibnamefont {Pereira}},\ and\
  \bibinfo {author} {\bibfnamefont {F.~B.}\ \bibnamefont {Ramos}},\ }\bibfield
  {title} {\bibinfo {title} {{Quantum spin circulator in Y junctions of
  Heisenberg chains}},\ }\href {https://doi.org/10.1103/PhysRevB.97.220402}
  {\bibfield  {journal} {\bibinfo  {journal} {Phys. Rev. B}\ }\textbf {\bibinfo
  {volume} {97}},\ \bibinfo {pages} {220402(R)} (\bibinfo {year}
  {2018})}\BibitemShut {NoStop}%
\bibitem [{\citenamefont {Buccheri}\ \emph {et~al.}(2019)\citenamefont
  {Buccheri}, \citenamefont {Egger}, \citenamefont {{G. Pereira}},\ and\
  \citenamefont {{B. Ramos}}}]{Buccheri2019}%
  \BibitemOpen
  \bibfield  {author} {\bibinfo {author} {\bibfnamefont {F.}~\bibnamefont
  {Buccheri}}, \bibinfo {author} {\bibfnamefont {R.}~\bibnamefont {Egger}},
  \bibinfo {author} {\bibfnamefont {R.}~\bibnamefont {{G. Pereira}}},\ and\
  \bibinfo {author} {\bibfnamefont {F.}~\bibnamefont {{B. Ramos}}},\ }\bibfield
   {title} {\bibinfo {title} {{Chiral Y junction of quantum spin chains}},\
  }\href {https://doi.org/https://doi.org/10.1016/j.nuclphysb.2019.03.005}
  {\bibfield  {journal} {\bibinfo  {journal} {Nucl. Phys. B}\ }\textbf
  {\bibinfo {volume} {941}},\ \bibinfo {pages} {794} (\bibinfo {year}
  {2019})}\BibitemShut {NoStop}%
\bibitem [{\citenamefont {{A. O. Gogolin, A. A. Nersesyan, and A. M.
  Tsvelik}}(1998)}]{Gogolin1998}%
  \BibitemOpen
  \bibfield  {author} {\bibinfo {author} {\bibnamefont {{A. O. Gogolin, A. A.
  Nersesyan, and A. M. Tsvelik}}},\ }\href@noop {} {\emph {\bibinfo {title}
  {{Bosonization and Strongly Correlated Systems}}}}\ (\bibinfo  {publisher}
  {Cambridge University Press},\ \bibinfo {address} {Cambridge},\ \bibinfo
  {year} {1998})\BibitemShut {NoStop}%
\bibitem [{\citenamefont {Affleck}\ and\ \citenamefont
  {Haldane}(1987)}]{Affleck1987}%
  \BibitemOpen
  \bibfield  {author} {\bibinfo {author} {\bibfnamefont {I.}~\bibnamefont
  {Affleck}}\ and\ \bibinfo {author} {\bibfnamefont {F.~D.~M.}\ \bibnamefont
  {Haldane}},\ }\bibfield  {title} {\bibinfo {title} {{Critical theory of
  quantum spin chains}},\ }\href {https://doi.org/10.1103/PhysRevB.36.5291}
  {\bibfield  {journal} {\bibinfo  {journal} {Phys. Rev. B}\ }\textbf {\bibinfo
  {volume} {36}},\ \bibinfo {pages} {5291} (\bibinfo {year}
  {1987})}\BibitemShut {NoStop}%
\bibitem [{\citenamefont {Nielsen}\ \emph {et~al.}(2011)\citenamefont
  {Nielsen}, \citenamefont {Cirac},\ and\ \citenamefont
  {Sierra}}]{Nielsen2011}%
  \BibitemOpen
  \bibfield  {author} {\bibinfo {author} {\bibfnamefont {A.~E.~B.}\
  \bibnamefont {Nielsen}}, \bibinfo {author} {\bibfnamefont {J.~I.}\
  \bibnamefont {Cirac}},\ and\ \bibinfo {author} {\bibfnamefont
  {G.}~\bibnamefont {Sierra}},\ }\bibfield  {title} {\bibinfo {title} {{Quantum
  spin Hamiltonians for the $SU(2)_k$ WZW model}},\ }\href
  {https://doi.org/10.1088/1742-5468/2011/11/p11014} {\bibfield  {journal}
  {\bibinfo  {journal} {J. Stat. Mech.: Theor. Exp.}\ }\textbf {\bibinfo
  {volume} {2011}},\ \bibinfo {pages} {P11014} (\bibinfo {year}
  {2011})}\BibitemShut {NoStop}%
\bibitem [{\citenamefont {Michaud}\ \emph {et~al.}(2012)\citenamefont
  {Michaud}, \citenamefont {Vernay}, \citenamefont {Manmana},\ and\
  \citenamefont {Mila}}]{Michaud2012}%
  \BibitemOpen
  \bibfield  {author} {\bibinfo {author} {\bibfnamefont {F.}~\bibnamefont
  {Michaud}}, \bibinfo {author} {\bibfnamefont {F.}~\bibnamefont {Vernay}},
  \bibinfo {author} {\bibfnamefont {S.~R.}\ \bibnamefont {Manmana}},\ and\
  \bibinfo {author} {\bibfnamefont {F.}~\bibnamefont {Mila}},\ }\bibfield
  {title} {\bibinfo {title} {{Antiferromagnetic Spin-$S$ Chains with Exactly
  Dimerized Ground States}},\ }\href
  {https://doi.org/10.1103/PhysRevLett.108.127202} {\bibfield  {journal}
  {\bibinfo  {journal} {Phys. Rev. Lett.}\ }\textbf {\bibinfo {volume} {108}},\
  \bibinfo {pages} {127202} (\bibinfo {year} {2012})}\BibitemShut {NoStop}%
\bibitem [{\citenamefont {Thomale}\ \emph {et~al.}(2012)\citenamefont
  {Thomale}, \citenamefont {Rachel}, \citenamefont {Schmitteckert},\ and\
  \citenamefont {Greiter}}]{Thomale2012}%
  \BibitemOpen
  \bibfield  {author} {\bibinfo {author} {\bibfnamefont {R.}~\bibnamefont
  {Thomale}}, \bibinfo {author} {\bibfnamefont {S.}~\bibnamefont {Rachel}},
  \bibinfo {author} {\bibfnamefont {P.}~\bibnamefont {Schmitteckert}},\ and\
  \bibinfo {author} {\bibfnamefont {M.}~\bibnamefont {Greiter}},\ }\bibfield
  {title} {\bibinfo {title} {{Family of spin-$S$ chain representations of
  SU(2)${}_{k}$ Wess-Zumino-Witten models}},\ }\href
  {https://doi.org/10.1103/PhysRevB.85.195149} {\bibfield  {journal} {\bibinfo
  {journal} {Phys. Rev. B}\ }\textbf {\bibinfo {volume} {85}},\ \bibinfo
  {pages} {195149} (\bibinfo {year} {2012})}\BibitemShut {NoStop}%
\bibitem [{\citenamefont {Michaud}\ \emph {et~al.}(2013)\citenamefont
  {Michaud}, \citenamefont {Manmana},\ and\ \citenamefont
  {Mila}}]{Michaud2013}%
  \BibitemOpen
  \bibfield  {author} {\bibinfo {author} {\bibfnamefont {F.}~\bibnamefont
  {Michaud}}, \bibinfo {author} {\bibfnamefont {S.~R.}\ \bibnamefont
  {Manmana}},\ and\ \bibinfo {author} {\bibfnamefont {F.}~\bibnamefont
  {Mila}},\ }\bibfield  {title} {\bibinfo {title} {{Realization of higher
  Wess-Zumino-Witten models in spin chains}},\ }\href
  {https://doi.org/10.1103/PhysRevB.87.140404} {\bibfield  {journal} {\bibinfo
  {journal} {Phys. Rev. B}\ }\textbf {\bibinfo {volume} {87}},\ \bibinfo
  {pages} {140404} (\bibinfo {year} {2013})}\BibitemShut {NoStop}%
\bibitem [{\citenamefont {Greiter}\ and\ \citenamefont
  {Thomale}(2009)}]{Greiter2009}%
  \BibitemOpen
  \bibfield  {author} {\bibinfo {author} {\bibfnamefont {M.}~\bibnamefont
  {Greiter}}\ and\ \bibinfo {author} {\bibfnamefont {R.}~\bibnamefont
  {Thomale}},\ }\bibfield  {title} {\bibinfo {title} {{Non-Abelian Statistics
  in a Quantum Antiferromagnet}},\ }\href
  {https://doi.org/10.1103/PhysRevLett.102.207203} {\bibfield  {journal}
  {\bibinfo  {journal} {Phys. Rev. Lett.}\ }\textbf {\bibinfo {volume} {102}},\
  \bibinfo {pages} {207203} (\bibinfo {year} {2009})}\BibitemShut {NoStop}%
\bibitem [{\citenamefont {Yao}\ and\ \citenamefont {Lee}(2011)}]{Yao2011}%
  \BibitemOpen
  \bibfield  {author} {\bibinfo {author} {\bibfnamefont {H.}~\bibnamefont
  {Yao}}\ and\ \bibinfo {author} {\bibfnamefont {D.-H.}\ \bibnamefont {Lee}},\
  }\bibfield  {title} {\bibinfo {title} {{Fermionic magnons, non-Abelian
  spinons, and spin quantum Hall effect from an exactly solvable spin-1/2
  Kitaev model with SU(2) symmetry}},\ }\href
  {https://doi.org/10.1103/PhysRevLett.107.087205} {\bibfield  {journal}
  {\bibinfo  {journal} {Phys. Rev. Lett.}\ }\textbf {\bibinfo {volume} {107}},\
  \bibinfo {pages} {087205} (\bibinfo {year} {2011})}\BibitemShut {NoStop}%
\bibitem [{\citenamefont {Tsvelik}(1990)}]{Tsvelik1990}%
  \BibitemOpen
  \bibfield  {author} {\bibinfo {author} {\bibfnamefont {A.~M.}\ \bibnamefont
  {Tsvelik}},\ }\bibfield  {title} {\bibinfo {title} {{Field-theory treatment
  of the Heisenberg spin-1 chain}},\ }\href
  {https://doi.org/10.1103/PhysRevB.42.10499} {\bibfield  {journal} {\bibinfo
  {journal} {Phys. Rev. B}\ }\textbf {\bibinfo {volume} {42}},\ \bibinfo
  {pages} {10499} (\bibinfo {year} {1990})}\BibitemShut {NoStop}%
\bibitem [{\citenamefont {Allen}\ and\ \citenamefont
  {S\'en\'echal}(2000)}]{Allen2000}%
  \BibitemOpen
  \bibfield  {author} {\bibinfo {author} {\bibfnamefont {D.}~\bibnamefont
  {Allen}}\ and\ \bibinfo {author} {\bibfnamefont {D.}~\bibnamefont
  {S\'en\'echal}},\ }\bibfield  {title} {\bibinfo {title} {{Spin-1 ladder: A
  bosonization study}},\ }\href {https://doi.org/10.1103/PhysRevB.61.12134}
  {\bibfield  {journal} {\bibinfo  {journal} {Phys. Rev. B}\ }\textbf {\bibinfo
  {volume} {61}},\ \bibinfo {pages} {12134} (\bibinfo {year}
  {2000})}\BibitemShut {NoStop}%
\bibitem [{\citenamefont {Chepiga}\ \emph
  {et~al.}(2016{\natexlab{a}})\citenamefont {Chepiga}, \citenamefont
  {Affleck},\ and\ \citenamefont {Mila}}]{Chepiga2016b}%
  \BibitemOpen
  \bibfield  {author} {\bibinfo {author} {\bibfnamefont {N.}~\bibnamefont
  {Chepiga}}, \bibinfo {author} {\bibfnamefont {I.}~\bibnamefont {Affleck}},\
  and\ \bibinfo {author} {\bibfnamefont {F.}~\bibnamefont {Mila}},\ }\bibfield
  {title} {\bibinfo {title} {Dimerization transitions in spin-1 chains},\
  }\href {https://doi.org/10.1103/PhysRevB.93.241108} {\bibfield  {journal}
  {\bibinfo  {journal} {Phys. Rev. B}\ }\textbf {\bibinfo {volume} {93}},\
  \bibinfo {pages} {241108} (\bibinfo {year} {2016}{\natexlab{a}})}\BibitemShut
  {NoStop}%
\bibitem [{\citenamefont {Yip}(2003)}]{Yip2003}%
  \BibitemOpen
  \bibfield  {author} {\bibinfo {author} {\bibfnamefont {S.~K.}\ \bibnamefont
  {Yip}},\ }\bibfield  {title} {\bibinfo {title} {{Dimer State of Spin-1 Bosons
  in an Optical Lattice}},\ }\href
  {https://doi.org/10.1103/PhysRevLett.90.250402} {\bibfield  {journal}
  {\bibinfo  {journal} {Phys. Rev. Lett.}\ }\textbf {\bibinfo {volume} {90}},\
  \bibinfo {pages} {250402} (\bibinfo {year} {2003})}\BibitemShut {NoStop}%
\bibitem [{\citenamefont {Garc\'{\i}a-Ripoll}\ \emph
  {et~al.}(2004)\citenamefont {Garc\'{\i}a-Ripoll}, \citenamefont
  {Martin-Delgado},\ and\ \citenamefont {Cirac}}]{Garcia2004}%
  \BibitemOpen
  \bibfield  {author} {\bibinfo {author} {\bibfnamefont {J.~J.}\ \bibnamefont
  {Garc\'{\i}a-Ripoll}}, \bibinfo {author} {\bibfnamefont {M.~A.}\ \bibnamefont
  {Martin-Delgado}},\ and\ \bibinfo {author} {\bibfnamefont {J.~I.}\
  \bibnamefont {Cirac}},\ }\bibfield  {title} {\bibinfo {title}
  {{Implementation of Spin Hamiltonians in Optical Lattices}},\ }\href
  {https://doi.org/10.1103/PhysRevLett.93.250405} {\bibfield  {journal}
  {\bibinfo  {journal} {Phys. Rev. Lett.}\ }\textbf {\bibinfo {volume} {93}},\
  \bibinfo {pages} {250405} (\bibinfo {year} {2004})}\BibitemShut {NoStop}%
\bibitem [{\citenamefont {Affleck}(1986)}]{Affleck1986}%
  \BibitemOpen
  \bibfield  {author} {\bibinfo {author} {\bibfnamefont {I.}~\bibnamefont
  {Affleck}},\ }\bibfield  {title} {\bibinfo {title} {{Exact critical exponents
  for quantum spin chains, non-linear $\sigma$-models at $\theta=\pi$ and the
  quantum Hall effect}},\ }\href {https://doi.org/10.1016/0550-3213(86)90167-7}
  {\bibfield  {journal} {\bibinfo  {journal} {Nucl. Phys. B}\ }\textbf
  {\bibinfo {volume} {265}},\ \bibinfo {pages} {409} (\bibinfo {year}
  {1986})}\BibitemShut {NoStop}%
\bibitem [{\citenamefont {L\"auchli}\ \emph {et~al.}(2006)\citenamefont
  {L\"auchli}, \citenamefont {Schmid},\ and\ \citenamefont
  {Trebst}}]{Lauchi2006}%
  \BibitemOpen
  \bibfield  {author} {\bibinfo {author} {\bibfnamefont {A.}~\bibnamefont
  {L\"auchli}}, \bibinfo {author} {\bibfnamefont {G.}~\bibnamefont {Schmid}},\
  and\ \bibinfo {author} {\bibfnamefont {S.}~\bibnamefont {Trebst}},\
  }\bibfield  {title} {\bibinfo {title} {{Spin nematics correlations in
  bilinear-biquadratic $S=1$ spin chains}},\ }\href
  {https://doi.org/10.1103/PhysRevB.74.144426} {\bibfield  {journal} {\bibinfo
  {journal} {Phys. Rev. B}\ }\textbf {\bibinfo {volume} {74}},\ \bibinfo
  {pages} {144426} (\bibinfo {year} {2006})}\BibitemShut {NoStop}%
\bibitem [{\citenamefont {Takhtajan}(1982)}]{Takhtajan1982}%
  \BibitemOpen
  \bibfield  {author} {\bibinfo {author} {\bibfnamefont {L.~A.}\ \bibnamefont
  {Takhtajan}},\ }\bibfield  {title} {\bibinfo {title} {{The picture of
  low-lying excitations in the isotropic Heisenberg chain of arbitrary
  spins}},\ }\href
  {https://doi.org/https://doi.org/10.1016/0375-9601(82)90764-2} {\bibfield
  {journal} {\bibinfo  {journal} {Phys. Lett. A}\ }\textbf {\bibinfo {volume}
  {87}},\ \bibinfo {pages} {479} (\bibinfo {year} {1982})}\BibitemShut
  {NoStop}%
\bibitem [{\citenamefont {Babujian}(1982)}]{Babujian1982}%
  \BibitemOpen
  \bibfield  {author} {\bibinfo {author} {\bibfnamefont {H.~M.}\ \bibnamefont
  {Babujian}},\ }\bibfield  {title} {\bibinfo {title} {{Exact solution of the
  one-dimensional isotropic Heisenberg chain with arbitrary spins $S$}},\
  }\href {https://doi.org/https://doi.org/10.1016/0375-9601(82)90403-0}
  {\bibfield  {journal} {\bibinfo  {journal} {Phys. Lett. A}\ }\textbf
  {\bibinfo {volume} {90}},\ \bibinfo {pages} {479} (\bibinfo {year}
  {1982})}\BibitemShut {NoStop}%
\bibitem [{\citenamefont {Affleck}\ \emph {et~al.}(1989)\citenamefont
  {Affleck}, \citenamefont {Gepner}, \citenamefont {Schulz},\ and\
  \citenamefont {Ziman}}]{Affleck1989}%
  \BibitemOpen
  \bibfield  {author} {\bibinfo {author} {\bibfnamefont {I.}~\bibnamefont
  {Affleck}}, \bibinfo {author} {\bibfnamefont {D.}~\bibnamefont {Gepner}},
  \bibinfo {author} {\bibfnamefont {H.~J.}\ \bibnamefont {Schulz}},\ and\
  \bibinfo {author} {\bibfnamefont {T.}~\bibnamefont {Ziman}},\ }\bibfield
  {title} {\bibinfo {title} {{Critical behaviour of spin-$s$ Heisenberg
  antiferromagnetic chains: analytic and numerical results}},\ }\href
  {https://doi.org/10.1088/0305-4470/22/5/015} {\bibfield  {journal} {\bibinfo
  {journal} {J. Phys. A: Math. Gen.}\ }\textbf {\bibinfo {volume} {22}},\
  \bibinfo {pages} {511} (\bibinfo {year} {1989})}\BibitemShut {NoStop}%
\bibitem [{\citenamefont {Chepiga}\ \emph
  {et~al.}(2016{\natexlab{b}})\citenamefont {Chepiga}, \citenamefont
  {Affleck},\ and\ \citenamefont {Mila}}]{Chepiga2016a}%
  \BibitemOpen
  \bibfield  {author} {\bibinfo {author} {\bibfnamefont {N.}~\bibnamefont
  {Chepiga}}, \bibinfo {author} {\bibfnamefont {I.}~\bibnamefont {Affleck}},\
  and\ \bibinfo {author} {\bibfnamefont {F.}~\bibnamefont {Mila}},\ }\bibfield
  {title} {\bibinfo {title} {{Comment on ``Frustration and multicriticality in
  the antiferromagnetic spin-1 chain''}},\ }\href
  {https://doi.org/10.1103/PhysRevB.94.136401} {\bibfield  {journal} {\bibinfo
  {journal} {Phys. Rev. B}\ }\textbf {\bibinfo {volume} {94}},\ \bibinfo
  {pages} {136401} (\bibinfo {year} {2016}{\natexlab{b}})}\BibitemShut
  {NoStop}%
\bibitem [{\citenamefont {Allen}\ and\ \citenamefont
  {S\'en\'echal}(1997)}]{Allen1997}%
  \BibitemOpen
  \bibfield  {author} {\bibinfo {author} {\bibfnamefont {D.}~\bibnamefont
  {Allen}}\ and\ \bibinfo {author} {\bibfnamefont {D.}~\bibnamefont
  {S\'en\'echal}},\ }\bibfield  {title} {\bibinfo {title} {{Non-Abelian
  bosonization of the frustrated antiferromagnetic spin-1/2 chain}},\ }\href
  {https://doi.org/10.1103/PhysRevB.55.299} {\bibfield  {journal} {\bibinfo
  {journal} {Phys. Rev. B}\ }\textbf {\bibinfo {volume} {55}},\ \bibinfo
  {pages} {299} (\bibinfo {year} {1997})}\BibitemShut {NoStop}%
\bibitem [{\citenamefont {Shelton}\ \emph {et~al.}(1996)\citenamefont
  {Shelton}, \citenamefont {Nersesyan},\ and\ \citenamefont
  {Tsvelik}}]{Shelton1996}%
  \BibitemOpen
  \bibfield  {author} {\bibinfo {author} {\bibfnamefont {D.~G.}\ \bibnamefont
  {Shelton}}, \bibinfo {author} {\bibfnamefont {A.~A.}\ \bibnamefont
  {Nersesyan}},\ and\ \bibinfo {author} {\bibfnamefont {A.~M.}\ \bibnamefont
  {Tsvelik}},\ }\bibfield  {title} {\bibinfo {title} {{Antiferromagnetic spin
  ladders: Crossover between spin $S=1/2$ and $S=1$ chains}},\ }\href
  {https://doi.org/10.1103/PhysRevB.53.8521} {\bibfield  {journal} {\bibinfo
  {journal} {Phys. Rev. B}\ }\textbf {\bibinfo {volume} {53}},\ \bibinfo
  {pages} {8521} (\bibinfo {year} {1996})}\BibitemShut {NoStop}%
\bibitem [{\citenamefont {Witten}(1984)}]{Witten1984}%
  \BibitemOpen
  \bibfield  {author} {\bibinfo {author} {\bibfnamefont {E.}~\bibnamefont
  {Witten}},\ }\bibfield  {title} {\bibinfo {title} {{Non-Abelian bosonization
  in two dimensions}},\ }\href {https://doi.org/10.1007/BF01215276} {\bibfield
  {journal} {\bibinfo  {journal} {Commun. Math. Phys.}\ }\textbf {\bibinfo
  {volume} {92}},\ \bibinfo {pages} {455} (\bibinfo {year} {1984})}\BibitemShut
  {NoStop}%
\bibitem [{\citenamefont {Affleck}\ \emph {et~al.}(1995)\citenamefont
  {Affleck}, \citenamefont {Ludwig},\ and\ \citenamefont
  {Jones}}]{AffleckJones1995}%
  \BibitemOpen
  \bibfield  {author} {\bibinfo {author} {\bibfnamefont {I.}~\bibnamefont
  {Affleck}}, \bibinfo {author} {\bibfnamefont {A.~W.~W.}\ \bibnamefont
  {Ludwig}},\ and\ \bibinfo {author} {\bibfnamefont {B.~A.}\ \bibnamefont
  {Jones}},\ }\bibfield  {title} {\bibinfo {title} {{Conformal-field-theory
  approach to the two-impurity Kondo problem: Comparison with numerical
  renormalization-group results}},\ }\href
  {https://doi.org/10.1103/PhysRevB.52.9528} {\bibfield  {journal} {\bibinfo
  {journal} {Phys. Rev. B}\ }\textbf {\bibinfo {volume} {52}},\ \bibinfo
  {pages} {9528} (\bibinfo {year} {1995})}\BibitemShut {NoStop}%
\bibitem [{\citenamefont {Zar\'and}\ \emph {et~al.}(2006)\citenamefont
  {Zar\'and}, \citenamefont {Chung}, \citenamefont {Simon},\ and\ \citenamefont
  {Vojta}}]{Zarand2006}%
  \BibitemOpen
  \bibfield  {author} {\bibinfo {author} {\bibfnamefont {G.}~\bibnamefont
  {Zar\'and}}, \bibinfo {author} {\bibfnamefont {C.-H.}\ \bibnamefont {Chung}},
  \bibinfo {author} {\bibfnamefont {P.}~\bibnamefont {Simon}},\ and\ \bibinfo
  {author} {\bibfnamefont {M.}~\bibnamefont {Vojta}},\ }\bibfield  {title}
  {\bibinfo {title} {{Quantum Criticality in a Double-Quantum-Dot System}},\
  }\href {https://doi.org/10.1103/PhysRevLett.97.166802} {\bibfield  {journal}
  {\bibinfo  {journal} {Phys. Rev. Lett.}\ }\textbf {\bibinfo {volume} {97}},\
  \bibinfo {pages} {166802} (\bibinfo {year} {2006})}\BibitemShut {NoStop}%
\bibitem [{\citenamefont {Rahmani}\ \emph {et~al.}(2012)\citenamefont
  {Rahmani}, \citenamefont {Hou}, \citenamefont {Feiguin}, \citenamefont
  {Oshikawa}, \citenamefont {Chamon},\ and\ \citenamefont
  {Affleck}}]{Rahmani2012}%
  \BibitemOpen
  \bibfield  {author} {\bibinfo {author} {\bibfnamefont {A.}~\bibnamefont
  {Rahmani}}, \bibinfo {author} {\bibfnamefont {C.-Y.}\ \bibnamefont {Hou}},
  \bibinfo {author} {\bibfnamefont {A.}~\bibnamefont {Feiguin}}, \bibinfo
  {author} {\bibfnamefont {M.}~\bibnamefont {Oshikawa}}, \bibinfo {author}
  {\bibfnamefont {C.}~\bibnamefont {Chamon}},\ and\ \bibinfo {author}
  {\bibfnamefont {I.}~\bibnamefont {Affleck}},\ }\bibfield  {title} {\bibinfo
  {title} {General method for calculating the universal conductance of strongly
  correlated junctions of multiple quantum wires},\ }\href
  {https://doi.org/10.1103/PhysRevB.85.045120} {\bibfield  {journal} {\bibinfo
  {journal} {Phys. Rev. B}\ }\textbf {\bibinfo {volume} {85}},\ \bibinfo
  {pages} {045120} (\bibinfo {year} {2012})}\BibitemShut {NoStop}%
\bibitem [{\citenamefont {Nayak}\ \emph {et~al.}(1999)\citenamefont {Nayak},
  \citenamefont {Fisher}, \citenamefont {Ludwig},\ and\ \citenamefont
  {Lin}}]{Nayak1999}%
  \BibitemOpen
  \bibfield  {author} {\bibinfo {author} {\bibfnamefont {C.}~\bibnamefont
  {Nayak}}, \bibinfo {author} {\bibfnamefont {M.~P.~A.}\ \bibnamefont
  {Fisher}}, \bibinfo {author} {\bibfnamefont {A.~W.~W.}\ \bibnamefont
  {Ludwig}},\ and\ \bibinfo {author} {\bibfnamefont {H.~H.}\ \bibnamefont
  {Lin}},\ }\bibfield  {title} {\bibinfo {title} {{Resonant multilead
  point-contact tunneling}},\ }\href
  {https://doi.org/10.1103/PhysRevB.59.15694} {\bibfield  {journal} {\bibinfo
  {journal} {Phys. Rev. B}\ }\textbf {\bibinfo {volume} {59}},\ \bibinfo
  {pages} {15694} (\bibinfo {year} {1999})}\BibitemShut {NoStop}%
\bibitem [{\citenamefont {Mahan}(2000)}]{mahan1990many}%
  \BibitemOpen
  \bibfield  {author} {\bibinfo {author} {\bibfnamefont {G.~D.}\ \bibnamefont
  {Mahan}},\ }\href {https://doi.org/https://doi.org/10.1007/978-1-4757-5714-9}
  {\emph {\bibinfo {title} {Many-Particle Physics}}}\ (\bibinfo  {publisher}
  {Springer New York, NY},\ \bibinfo {year} {2000})\BibitemShut {NoStop}%
\bibitem [{\citenamefont {Guo}\ and\ \citenamefont {White}(2006)}]{Guo2006}%
  \BibitemOpen
  \bibfield  {author} {\bibinfo {author} {\bibfnamefont {H.}~\bibnamefont
  {Guo}}\ and\ \bibinfo {author} {\bibfnamefont {S.~R.}\ \bibnamefont
  {White}},\ }\bibfield  {title} {\bibinfo {title} {{Density matrix
  renormalization group algorithms for Y-junctions}},\ }\href
  {https://doi.org/10.1103/PhysRevB.74.060401} {\bibfield  {journal} {\bibinfo
  {journal} {Phys. Rev. B}\ }\textbf {\bibinfo {volume} {74}},\ \bibinfo
  {pages} {060401} (\bibinfo {year} {2006})}\BibitemShut {NoStop}%
\bibitem [{\citenamefont {Rahmani}\ \emph {et~al.}(2010)\citenamefont
  {Rahmani}, \citenamefont {Hou}, \citenamefont {Feiguin}, \citenamefont
  {Chamon},\ and\ \citenamefont {Affleck}}]{Rahmani2010}%
  \BibitemOpen
  \bibfield  {author} {\bibinfo {author} {\bibfnamefont {A.}~\bibnamefont
  {Rahmani}}, \bibinfo {author} {\bibfnamefont {C.-Y.}\ \bibnamefont {Hou}},
  \bibinfo {author} {\bibfnamefont {A.}~\bibnamefont {Feiguin}}, \bibinfo
  {author} {\bibfnamefont {C.}~\bibnamefont {Chamon}},\ and\ \bibinfo {author}
  {\bibfnamefont {I.}~\bibnamefont {Affleck}},\ }\bibfield  {title} {\bibinfo
  {title} {{How to find conductance tensors of quantum multiwire junctions
  through static calculations: Application to an interacting $Y$ junction}},\
  }\href {https://doi.org/10.1103/PhysRevLett.105.226803} {\bibfield  {journal}
  {\bibinfo  {journal} {Phys. Rev. Lett.}\ }\textbf {\bibinfo {volume} {105}},\
  \bibinfo {pages} {226803} (\bibinfo {year} {2010})}\BibitemShut {NoStop}%
\bibitem [{\citenamefont {Kalmeyer}\ and\ \citenamefont
  {Laughlin}(1987)}]{Kalmeyer1987}%
  \BibitemOpen
  \bibfield  {author} {\bibinfo {author} {\bibfnamefont {V.}~\bibnamefont
  {Kalmeyer}}\ and\ \bibinfo {author} {\bibfnamefont {R.~B.}\ \bibnamefont
  {Laughlin}},\ }\bibfield  {title} {\bibinfo {title} {Equivalence of the
  resonating-valence-bond and fractional quantum hall states},\ }\href
  {https://doi.org/10.1103/PhysRevLett.59.2095} {\bibfield  {journal} {\bibinfo
   {journal} {Phys. Rev. Lett.}\ }\textbf {\bibinfo {volume} {59}},\ \bibinfo
  {pages} {2095} (\bibinfo {year} {1987})}\BibitemShut {NoStop}%
\bibitem [{\citenamefont {Mussardo}(2010)}]{Mussardo2010}%
  \BibitemOpen
  \bibfield  {author} {\bibinfo {author} {\bibfnamefont {G.}~\bibnamefont
  {Mussardo}},\ }\href {https://doi.org/10.1093/oso/9780198788102.001.0001}
  {\emph {\bibinfo {title} {Statistical Field Theory: An Introduction to
  Exactly Solved Models in Statistical Physics}}},\ \bibinfo {edition} {1st}\
  ed.\ (\bibinfo  {publisher} {Oxford University Press},\ \bibinfo {address}
  {New York},\ \bibinfo {year} {2010})\BibitemShut {NoStop}%
\bibitem [{\citenamefont {{P. Di Francesco, P. Mathieu and D.
  S\'{e}n\'{e}chal}}(1997)}]{DiFrancesco1997}%
  \BibitemOpen
  \bibfield  {author} {\bibinfo {author} {\bibnamefont {{P. Di Francesco, P.
  Mathieu and D. S\'{e}n\'{e}chal}}},\ }\href
  {https://doi.org/10.1007/978-1-4612-2256-9} {\emph {\bibinfo {title}
  {{Conformal Field Theory}}}},\ Graduate Texts in Contemporary Physics\
  (\bibinfo  {publisher} {Springer},\ \bibinfo {address} {New York},\ \bibinfo
  {year} {1997})\BibitemShut {NoStop}%
\bibitem [{\citenamefont {Haldane}(1988)}]{Haldane1988}%
  \BibitemOpen
  \bibfield  {author} {\bibinfo {author} {\bibfnamefont {F.~D.~M.}\
  \bibnamefont {Haldane}},\ }\bibfield  {title} {\bibinfo {title} {{Model for a
  Quantum Hall Effect without Landau Levels: Condensed-Matter Realization of
  the ``Parity Anomaly''}},\ }\href
  {https://doi.org/10.1103/PhysRevLett.61.2015} {\bibfield  {journal} {\bibinfo
   {journal} {Phys. Rev. Lett.}\ }\textbf {\bibinfo {volume} {61}},\ \bibinfo
  {pages} {2015} (\bibinfo {year} {1988})}\BibitemShut {NoStop}%
\end{thebibliography}%

%%%%%%%%%%%%%%%%%%%%%%%%%%%%%%%%%%%%%%%%%%%%%%%%%%%%%%%%%%

\end{document}